\documentclass[aps,prd,reprint,nofootinbib]{revtex4-2}

\usepackage{float}
\usepackage{natbib}
\usepackage{amsmath}
\usepackage{graphicx}
\usepackage{multirow}
\usepackage[colorlinks=true, allcolors=blue]{hyperref}
\usepackage{longtable}
\usepackage{booktabs}
\usepackage{subfigure}

\newcommand{\rev}[1]{\textcolor{black}{#1}}

\begin{document}


\title{Counting voids and filaments: 
Betti Curves as a Topological Probe of Cosmology}


\author{Jiayi Li}
\email{l-jy21@mails.tsinghua.edu.cn}
\affiliation{Department of Astronomy, Tsinghua University, Beijing 100084, China}
\author{Cheng Zhao}
\email{czhao@tsinghua.edu.cn}

\affiliation{Department of Astronomy, Tsinghua University, Beijing 100084, China}


\date{\today}

\begin{abstract}
Topological analysis of galaxy distributions has gathered increasing attention in cosmology, as they are able to capture non-Gaussian features of large-scale structures (LSS) that are overlooked by conventional two-point clustering statistics. We utilize Betti curves, a summary statistic derived from persistent homology, to characterize the multiscale topological features of the LSS, including connected components, loops, and voids, as a complementary cosmological probe. \rev{Using halo catalogs from the \textsc{Quijote} suite, we develop a coherent and extendable analysis framework that combines Betti-curve measurements, automated machine-learning-based emulators, and Bayesian inference for cosmological parameter estimation. Within this framework, we assess the sensitivity of Betti curves to cosmological parameters and train emulators to model their dependence on cosmology.} Our Bayesian inference recovers unbiased estimation of cosmological parameters, notably $n_{\mathrm{s}}$, $\sigma_8$, and $\Omega_{\mathrm{m}}$, while validation on sub-box simulations confirms robustness against cosmic variance. We further investigate the impact of redshift-space distortions (RSD) on Betti curves and demonstrate that including RSD enhances sensitivity to growth-related parameters. \rev{When combined with the power spectrum, Betti curves break parameter degeneracies and lead to significantly tighter joint constraints than the power spectrum alone on parameters such as $n_{\mathrm{s}}$, $\sigma_8$, and $w$. These results establish Betti curves as a viable and complementary topological observable for cosmological parameter inference, and motivate further theoretical and observational developments toward their application in future galaxy surveys.}
\end{abstract}


\maketitle

\section{\label{sec:introduction}Introduction}
The large-scale structure (LSS) of the Universe encodes critical information about the cosmic composition, expansion, and evolution\cite{LSS1985, LSS1980}. Spectroscopic surveys such as 2-degree Field Galaxy Redshift Survey (2dFGRS \footnote{\url{http://www.2dfgrs.net/}}; \cite{2df}), Sloan Digital Sky Survey, Baryon Oscillation Spectroscopic Survey and extended Baryon Oscillation Spectroscopic Survey (SDSS, BOSS and eBOSS \footnote{\url{https://www.sdss.org/}, \url{https://www.sdss4.org/surveys/boss/},
\url{https://www.sdss4.org/surveys/eboss/}}; \cite{SDSS,BOSS,eBOSS}), and the ongoing survey Dark Energy Spectroscopic Instrument (DESI \footnote{\url{https://www.desi.lbl.gov/}}; \cite{DESI}), have collected hundreds of thousands to tens of millions of spectra, created a precise 3D map up to $z\sim3$, revealing the cosmic web structures \cite{CosmicWeb1996,CosmicWeb14}. 
With current observational data, the standard cosmological parameters have been constrained to percent-level \cite{DESIConstraint}, ushering in the era of precise cosmology. Notably, the most recent results from the DESI collaboration suggest a possible hint of dynamic dark energy \cite{DESIDDE}. Meanwhile, numerous compelling questions beyond the standard model -- such as Primordial Non-Gaussianity (PNG) and Modified Gravity (MG) theories -- await further investigation with improved data.
Upcoming stage-V surveys, including MUltiplexed Survey Telescope (MUST \footnote{\url{https://must.astro.tsinghua.edu.cn/}}; \cite{MUST}), Stage-5 Spectroscopic Experiment (Spec-S5 \footnote{\url{https://spec-s5.org/}}; \cite{SpecS}), and Wide-field Spectroscopic Telescope (WST \footnote{\url{https://www.wstelescope.com/}}; \cite{WST}), will further expand the survey coverage. These missions will extend redshift reach to $z\sim 5$ and collect over 100 million galaxy redshifts. As a result, Stage-V surveys are expected to tighten constraints on standard cosmological parameters to sub-percent-level and improve sensitivity to parameters characterizing PNG and MG \cite{MUST,SpecS,WST}.

To address these emerging challenges and opportunities, it is crucial to reassess the tools used in cosmological analyses. 
Conventional cosmological analyses (i.e., Baryon Acoustic Oscillations (BAO) \cite{BAO}, Redshift Space Distortion (RSD) \cite{RSD} measurements) are mostly based on two-point clustering statistics, such as two-point correlation function and its Fourier transform, the power spectrum \cite{BOSS2pcf22,BOSSConstraint,DESConstraint,DESIFullshape}. These two-point statistics fully capture the information of a Gaussian random field, so it is important for cosmological studies since the primordial density field is well described by a nearly Gaussian field with small amplitude fluctuations as inferred from Cosmic Microwave Background (CMB) \cite{CMB,PlanckConstraint,CMBmap}. 
 However, due to the nonlinear evolution of structures, non-Gaussian features become prominent on small scales and low redshift, two-point statistics are not able to fully capture the information from the spectroscopic data. As current and upcoming surveys continue to expand in volume and accumulate data, the statistical precision of high-order clustering patterns will be high enough to provide sufficient cosmological information.
 
To fully exploit this wealth of information, it is therefore essential to develop alternative clustering statistics that can probe the non-Gaussian features of LSS, thereby improving the constraints of the cosmological parameters \cite{HigherOrderStat23,HigherOrderStat25}. 
The natural extension of two-point statistics is $N$-point statistics, where the lowest-order cases -- 3-point correlation function (3PCF) or its Fourier transform, bispectrum -- has been extensively studied \cite{Bispectrum00,Bispectrum04,Bispectrum23}. In addition to $N$-point statistics, alternative approaches include Void Size Function (VSF) \cite{VSF25}, nearest neighbor distribution (NN) \cite{NNdistribution20}, one-point Probability Distribution Function (PDF, also known as counts-in-cells statistics) \cite{PDF20}, Minkowski Functionals (MF) \cite{MF1994}, and many others.
Compared to the conventional 2PCF, these statistics have the potential to improve cosmological parameter constraints, help break parameter degeneracies, and offer greater sensitivity to scientific cases beyond standard cosmology, such as neutrino mass, primordial non-Gaussianity, and modified gravity \cite{PDF20,3PCF-png1994,3PCF-png24,MF-MG24,MG18,Void-MG19}.
Moreover, it is possible to integrate two-point clustering measurements and higher-order statistics to maximize cosmological information \cite{Combine21,Combine23}.  
Nonetheless, many of these point statistics -- bispectrum, NN, and PDF -- encounter computational challenges or are limited in capturing global information. Notably, the VSF, though geometrically motivated, remains fundamentally a point statistic and inherits similar limitations.
The genus statistic, Minkowski functionals, captures morphological information but are inherently non-local, posing challenges in overlapping or percolating structures \cite{MF20}. 

In light of these challenges, a method from Topological Data Analysis (TDA) -- specifically Persistent Homology (PH) -- quantifies the multiscale topology of data, offering a powerful and complementary perspective for analyzing complex structures (see \cite{TDA00,TDA16,TDA21} for a review).
The rich framework of PH has been successfully applied in various fields, such as computer vision and computer graphics \cite{TDA-cv14,TDA-cv10,TDA-cv09},
systems biology \cite{TDA-bio20}, materials science \cite{TDA-material16,TDA-material15},  complex systems and chaotic dynamics \cite{TDA-ComplexSystem19,TDA-chaotic16}.
In cosmology, PH enables the characterization of connected components, loops, and voids in the cosmic web, thereby providing insights into gravitational collapse and the cosmic expansion history. It has been used for the detection of BAO signal \cite{TDA-BAO20}, distinguishing dark energy models \cite{TDA-DE11}, measurements of structure growth and intrinsic alignment \cite{TDA-DES22}, as well as identification and evolution of cosmic web structures \cite{TDA-cosmicWeb19,TDA-cosmicWeb21}. 
Moreover, it has been found that PH can constrain standard cosmological parameters and primordial non-Gaussianity through Fisher forecast \cite{TDA-Mnu24,Cosmo_PH_Fisher}. \cite{PDCNN24} constructs a Convolutional Neural Network (CNN) for the Persistence Diagram (PD), the direct output of PH, and constrains cosmological parameters within the Bayesian framework using simulated halo catalogs. The study demonstrates that PD is a promising tool for cosmological parameter inference. However, as a 2-D field-level statistic, \rev{PD requires modeling a higher-dimensional representation and may be more sensitive to cosmic variance and systematic effects in the simulated training set.}

\rev{The primary goal of this work is to provide a proof-of-concept demonstration and to establish a coherent emulator-based cosmological inference pipeline for a functional summaries of PD, Betti curves.}
Betti curves transform PD into smooth, interpretable functions that are easier to model than PD. 
Using halo catalogs from cosmological simulations, we construct Gaussian process emulators using the automated machine learning (AutoML) technique for Betti curves, enabling efficient and scalable extraction of cosmological information. With these trained emulators, we demonstrate how Betti curves can constrain standard cosmological parameters, as well as extensions to the standard model, including the total neutrino mass and the dark energy equation-of-state parameter. We further investigate the impact of RSD on Betti curves and their corresponding parameter constraints, offering a more realistic and comprehensive assessment of the method's robustness under observational effects. Lastly, we compare the parameter constraints obtained from Betti curves with those derived from the power spectrum, providing insight into the complementary nature of the topological and point clustering information encoded in large-scale structure statistics. 

The paper is structured as follows. Section~\ref{sec:BC from sim} introduces the basis of Betti curves, outlines the dataset used in this work and assesses the sensitivity of Betti curves to cosmological parameters. Section~\ref{sec:emulator} describes the construction of data vectors, the training of the emulator, and the validation of the emulator.
Section~\ref{sec:result} presents the results of parameter constraints through Betti curves and power spectrum, analyzes the effects of RSD on parameter constraints, and demonstrates the statistical stability of our pipeline. Finally, Section~\ref{sec:conclusions} interprets our findings and discusses future directions. 

\section{Betti curve measurements from simulations  \label{sec:BC from sim}}
\subsection{Basis of persistent homology \label{sec:PH}}
Persistent homology is a technique for extracting topological information from data, whether a point cloud or a continuous field \cite{TDA21, TDA16}. It analyzes the shape of data across multiple scales, capturing the appearance and disappearance of topological features such as connected components, loops, and voids through a filtration process. The tracked features can be summarized into topological statistics that characterize the underlying topology of the hierarchical cosmic web and can be used to constrain cosmology. This paper specifically focuses on the persistent homology of point cloud data, such as halo and galaxy catalogs.

To extract the underlying topology from LSS, the observational data must be represented in a manner that makes its topological structure explicit. A useful representation is the simplicial complex, a combinatorial structure composed of simplices that are systematically connected--vertex to vertex, edge to edge, and face to face. Formally, a $k$-simplex is the convex hull of $k+1$ affinely independent points. For instance, in three-dimensional space, the fundamental simplices include: a 0-simplex (a point), a 1-simplex (an edge), a 2-simplex (a triangle), and a 3-simplex (a tetrahedron). A simplicial complex is a collection of connected simplices, with the requirement that any face of a simplex within the complex is also included in the complex. In LSS, dark matter halos or galaxies are treated as 0-simplices, while their spatial correlations define higher-dimensional simplices. Consequently, the simplicial complex is a natural representation that reveals the underlying topology of the cosmic web.

\begin{figure*}
    \centering
    \includegraphics[width=1\linewidth]{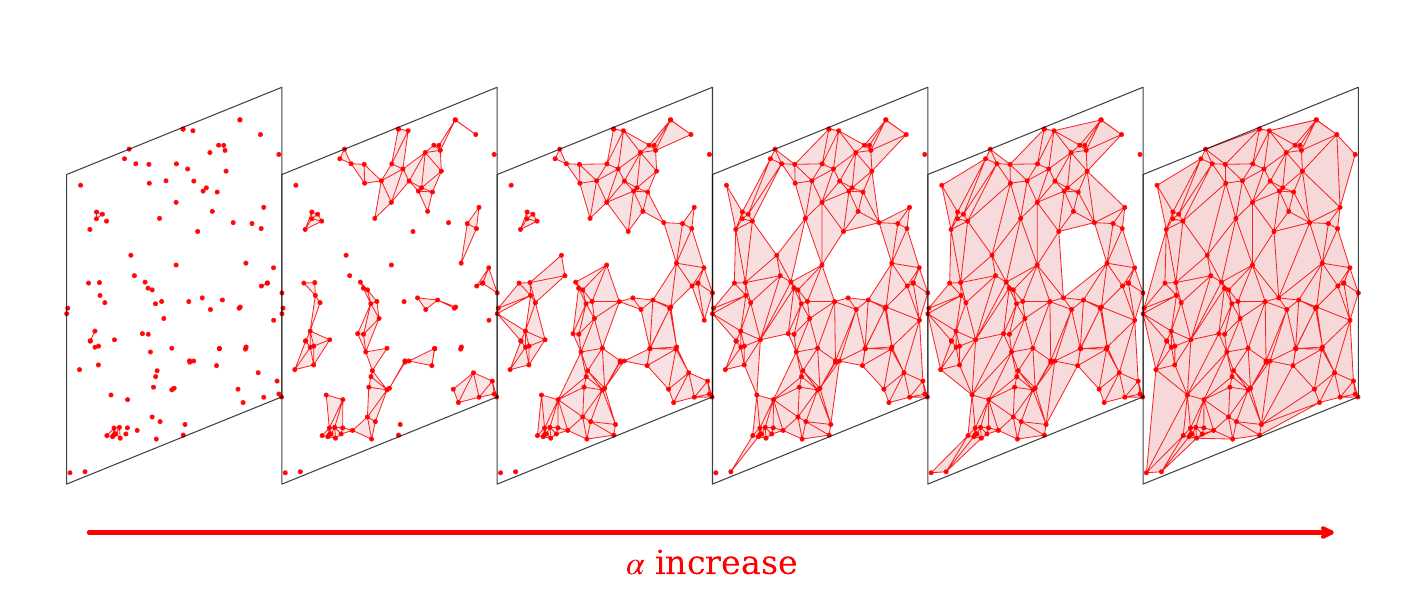}
    \caption{Visualization of alpha filtration.}
    \label{fig:alpha filtration}
\end{figure*}

For the specific simplicial complex constructed from data, this study focuses on the alpha complex \cite{AlphaShape}, which is the subcomplex of Delaunay triangulation. Given a scale parameter $\alpha>0$, the alpha complex consists of all simplices from the Delaunay triangulation whose minimum circumscribing sphere has a radius smaller than $\sqrt{\alpha}$. An alpha filtration is a nested sequence of alpha complexes parameterized by the filtration value $\alpha$, allowing for the characterization of the multiscale topology of the cosmic web: small $\alpha$ resolves isolated halos, while larger $\alpha$ reveals filaments and voids. At $\alpha=0$, the alpha complex consists solely of the individual data points. As $\alpha$ increases, discrete vertices connect to form filaments, loops, and voids. With increasing $\alpha$, smaller structures progressively merge into larger ones as visualized in Figure~\ref{fig:alpha filtration}. 
In the limit of $\alpha \to \infty$, the alpha complex converges to the Delaunay triangulation. 
The filtration process provides rich information about the topological features of different dimensions in data across multiple scales. Specifically, 0-dimensional features correspond to connected components, 1-dimensional features correspond to loops, and 2-dimensional features correspond to voids.

The direct output of the alpha filtration process is the persistence diagram, which records the birth and death filtration values of all topological features that emerge throughout the process.
While the persistence diagram provides a comprehensive visualization of the filtration results, it is not well-suited for statistical analysis due to its complex structure. To facilitate statistical analysis, one can derive functional summaries from the persistence diagram by sacrificing some information in exchange for a more tractable representation. 
Several functional summaries have been introduced in the literature, such as Betti curve \cite{BC11}, silhouettes \cite{Silhouette}, and entropy summary functions \cite{Entropy20}. The work focuses on the Betti curve, as it offers a particularly intuitive physical interpretation. A Betti curve plots Betti number $\beta_k(\alpha)$ as a function of filtration value $\alpha$. The Betti number $\beta_k(\alpha)$ summarizes the topology of the cosmic web at scale $\alpha$, counting the \textit{k}-dimensional features: 
\begin{itemize}
    \item $\beta_0(\alpha)$: Number of connected components,
    \item $\beta_1(\alpha)$: Number of independent loops,
    \item $\beta_2(\alpha)$: Number of voids \footnote{\rev{We note that the “voids” in our context are not underdense regions in common sense, but rather hollow structures bounded by 2‑simplices (triangles). At small filtration values, high‑density regions can also generate numerous such voids.}}.
\end{itemize}
\rev{As a compressed representation of PD, Betti curves trade part of the detailed pairing information for improved interpretability, smoothness, and computational tractability. Although this compression discards explicit birth–death correspondences, persistence information is implicitly encoded in the scale dependence of the curves: long-lived features contribute over broader filtration ranges than short-lived ones. This makes Betti curves particularly suitable for emulator-based cosmological inference while retaining sensitivity to large-scale topological structure.} 

\subsection{Simulation \label{sec:simulation}}
To develop and validate a pipeline for constraining cosmology using Betti curves, we use simulated halo catalogs from the $N$-body simulations in the \textsc{QUIJOTE} suite \cite{Quijote_sims}. Each simulation follows the evolution of $512^3$ particles in a periodic box of length $1\  (\text{Gpc}/h)$. The simulations are run using the TreePM code \textsc{GADGET}-III with initial conditions (ICs) generated at redshift $z=127$ using either second-order perturbation theory (2LPT) or Zeldovich approximation (ZA). Halos are identified using the Friends-of-Friends(FoF) algorithm. In this work, we analyze halo catalogs at $z=0.5$, a redshift comparable to that of galaxies observed in surveys such as the BOSS and DESI sample \cite{BOSS-DR12,DESI-DR1}, and also to be comparable with forecast in \cite{Cosmo_PH_Fisher}. 

To introduce the RSD effect into our simulated datasets, we map the real-space positions of halos $\boldsymbol{x}_{\text{real}}$ to their corresponding redshift-space positions $\boldsymbol{x}_{\text{redshift}}$ using the relation below:
\rev{
\begin{equation}
\label{eq:RSD}
\boldsymbol{x}_{\text{redshift}} = \boldsymbol{x}_{\text{real}} + \frac{\boldsymbol{v}\cdot \hat{\boldsymbol{n}}}{a(z=0.5)H(z=0.5)}\hat{\boldsymbol{n}},
\end{equation}}
Without loss of generality, we take the line-of-sight direction to be $\hat{\boldsymbol{n}} = (0,0,1)$. \rev{This implementation is intended as a controlled first step to assess the intrinsic response of Betti curves to velocity distortions, while avoiding additional complications introduced by survey geometry. Extending Betti-curve analyses to full survey geometries will be an essential step toward robust observational applications.}

To assess the sensitivity of Betti curves to cosmological parameters and validate the emulator, we employ both fiducial simulations in agreement with Planck's latest constraints \cite{PlanckConstraint} and a large number of comparable realizations with various cosmological parameters. 
For fiducial cosmology, we use two subsets: fiducial$\_$ZA and fiducial. The former refers to the fiducial simulations with ICs generated using ZA, the latter refers to those with ICs generated using 2LPT.
For the individual parameter variations (named as $\theta\_$m, $\theta\_$p, or $\theta\_$pp), the parameters vary from the fiducial cosmology by $\Delta \theta$, where $\{\Delta \Omega_\text{m}, \Delta \Omega_\text{b},\Delta h, \Delta n_{\text{s}},\Delta \sigma_{\text{8}},\Delta w\}=
\{\pm 0.01, \pm 0.002, \pm 0.02,\pm 0.02,\pm 0.015,\pm 0.05\}$
\footnote{For the simulations with $w$ variations and non-zero neutrino mass, the initial conditions are generated using ZA, while others using 2LPT. Thus, we compare the fiducial$\_$ZA subset with $w$ variations and non-zero neutrino mass, and the fiducial subset with others, respectively, when comparing the Betti curves under different cosmologies.}.
And the total mass of neutrinos $M_{\nu}$ takes values $0.1\  \text{and}\ 0.2$ eV. For each cosmology, we use 500 realizations to quantify the sensitivity of Betti curves to cosmological parameters.
The training set used for numerical modeling is the nwLH subset with various $\{ \Omega_\text{m},  \Omega_\text{b}, h,  n_{\text{s}}, \sigma_{\text{8}},w,M_{\nu}\}$. The nwLH set contains 2000 cosmologies, each with a single realization, and all initial conditions are generated using ZA. A summary of all simulations used in this work is provided in Table~\ref{tab:simulation}.

\begin{table*}
    \centering
    \begin{tabular}{lccccccccc}
        \toprule
        Category & $\Omega_\text{m}$ & $\Omega_\text{b}$ & $h$ & $n_\text{s}$ & $\sigma_8$ & $M_{\nu}$~(eV) & $w$ & Realizations & ICs \\
        \midrule
        \multicolumn{10}{c}{\textbf{Fiducial simulations}} \\
        \midrule
        fiducial & 0.3175 & 0.049 & 0.6711 & 0.9624 & 0.834 & 0.0 & -1 & 500 & 2LPT \\
        fiducial\_ZA & 0.3175 & 0.049 & 0.6711 & 0.9624 & 0.834 & 0.0 & -1 & 500 & ZA \\
        \midrule
        \multicolumn{10}{c}{\textbf{Variations around fiducial}} \\
        \midrule
        Om\_p & 0.3275 & 0.049 & 0.6711 & 0.9624 & 0.834 & 0.0 & -1 & 500 & 2LPT \\
        Om\_m & 0.3075 & 0.049 & 0.6711 & 0.9624 & 0.834 & 0.0 & -1 & 500 & 2LPT \\
        Ob\_p & 0.3175 & 0.051 & 0.6711 & 0.9624 & 0.834 & 0.0 & -1 & 500 & 2LPT \\
        Ob\_m & 0.3175 & 0.047 & 0.6711 & 0.9624 & 0.834 & 0.0 & -1 & 500 & 2LPT \\
        h\_p  & 0.3175 & 0.049 & 0.6911 & 0.9624 & 0.834 & 0.0 & -1 & 500 & 2LPT \\
        h\_m  & 0.3175 & 0.049 & 0.6511 & 0.9624 & 0.834 & 0.0 & -1 & 500 & 2LPT \\
        ns\_p & 0.3175 & 0.049 & 0.6711 & 0.9824 & 0.834 & 0.0 & -1 & 500 & 2LPT \\
        ns\_m & 0.3175 & 0.049 & 0.6711 & 0.9424 & 0.834 & 0.0 & -1 & 500 & 2LPT \\
        s8\_p & 0.3175 & 0.049 & 0.6711 & 0.9624 & 0.849 & 0.0 & -1 & 500 & 2LPT \\
        s8\_m & 0.3175 & 0.049 & 0.6711 & 0.9624 & 0.819 & 0.0 & -1 & 500 & 2LPT \\
        w\_p  & 0.3175 & 0.049 & 0.6711 & 0.9624 & 0.834 & 0.0 & -0.95 & 500 & ZA \\
        w\_m  & 0.3175 & 0.049 & 0.6711 & 0.9624 & 0.834 & 0.0 & -1.05 & 500 & ZA \\
        Mnu\_p & 0.3175 & 0.049 & 0.6711 & 0.9624 & 0.834 & 0.1 & -1 & 500 & ZA \\
        Mnu\_pp & 0.3175 & 0.049 & 0.6711 & 0.9624 & 0.834 & 0.2 & -1 & 500 & ZA \\
        \midrule
        \multicolumn{10}{c}{\textbf{Latin Hypercube (nwLH)}} \\
        \midrule
        nwLH & [0.1, 0.5] & [0.03, 0.07] & [0.5, 0.9] & [0.8, 1.2] & [0.6, 1.0] & [0.01, 1] & [-1.3, -0.7] & 2000 & ZA \\
        \bottomrule
    \end{tabular}
    \caption{Simulation sets from \textsc{QUIJOTE} used in this work}
    \label{tab:simulation}
\end{table*}

\subsection{From halo catalog to Betti curve \label{sec:normalize}}
We use the \texttt{GUDHI}\footnote{\texttt{GUDHI} is a \texttt{C++} library with a \texttt{Python} interface for Topological Data Analysis, offering data structures and algorithms to construct simplicial complexes and compute persistent homology, see \url{https://gudhi.inria.fr/}} library \cite{gudhi:urm, gudhi:AlphaComplex, gudhi:PersistenceRepresentations} to compute persistent homology and derive Betti curves of the simulated halo catalogs.
Since the halo catalogs are simulated in 3D periodic box, we apply periodic boundary conditions when computing persistent homology using the \texttt{alpha\_complex\_wrapper} code\footnote{See \url{https://github.com/ajouellette/alpha_complex_wrapper}} \cite{AlphaComplexWrapper}. 
As mentioned in section~\ref{sec:PH}, the filtration value $\alpha$ represents squared distances. To maintain consistency with the comoving scale, we replace the filtration value with its square root after computing persistent homology, ensuring that the adjusted filtration value ${\alpha}^{\prime}=\sqrt{\alpha}$ has the dimension of (Mpc$/h$). 
When comparing Betti curves across different halo catalogs, the curves shift along the ${\alpha}^{\prime}$-axis due to varying halo number densities \cite{BCrescale}. For example, the $\beta_1$ (loop) curve peaks at smaller filtration values in higher-density catalogs than in lower-density ones. To ensure comparability across catalogs, we normalize ${\alpha}^{\prime}$ by the halo number density as in \cite{BCrescale}, defining the dimensionless filtration value $\hat{\alpha}$ as:

\begin{equation}
\label{eq:alpha}
    \hat{\alpha} = \alpha^{\prime}/\ell,
\end{equation}

where $L$ is the simulation box size, $N$ is the total number of halos, and $\ell = L/N^{1/3}$ can be interpreted as the halo average separation.

\begin{figure}
    \includegraphics[width=\linewidth]{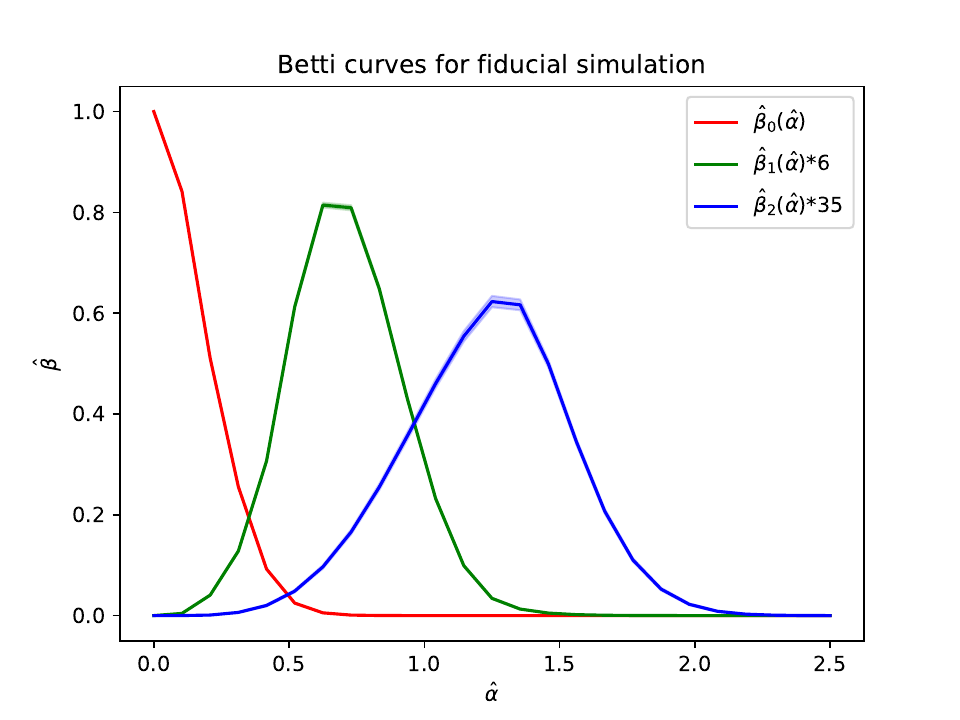}
    \caption{Normalized Betti curves for fiducial simulations in three dimensions. We rescale the amplitude of the Betti curves here for better visualization. The solid dark lines represent the mean Betti curves in three dimensions, while the shaded regions are the 1-$\sigma$ scatter of Betti curves inferred from 500 realizations.}
    \label{fig:Betti curves}
\end{figure}

Additionally, we normalize the Betti number $\beta$ to mitigate the influence of observation volume. The normalized Betti number is defined as:
\rev{
\begin{equation}
\label{eq:beta}
    \hat{\beta} = \beta\cdot \ell^3/L^3 = \beta/N.
\end{equation}
}
Here, the definition of $L$ and $\ell$ follows the equation~\ref{eq:alpha}. The division by $L^3$ accounts for the observation volume, while the multiplication by $\ell^3$ ensures the Betti number remains dimensionless. We compute the Betti curves in the range [0, 2.5], dividing the interval into 25 bins, which is sufficient to capture all relevant features. Beyond that range, where the scale is several times larger than $\ell$, all Betti curves vanish to zero in our simulations. This is because the persistent lifetime of a feature generally depends on the length of the longest edge forming the feature, which is around $\mathcal{O}(\ell)$. 
Figure~\ref{fig:Betti curves} presents the Betti curves computed from 500 fiducial simulations. 
The amplitude of each curve represents the number of topological structures (i.e., clusters, tunnels, and voids) at different scales.
The declining $\hat{\beta}_0(\hat{\alpha})$ reflects the hierarchical merging of clusters as $\hat\alpha$ increases. Meanwhile, the peaks in $\hat\beta_1(\hat\alpha)$ and $\hat\beta_2(\hat\alpha)$ correspond to prominent tunnels and voids that persist over large scales.

\begin{figure*}
    \centering
    \includegraphics[width=\linewidth]{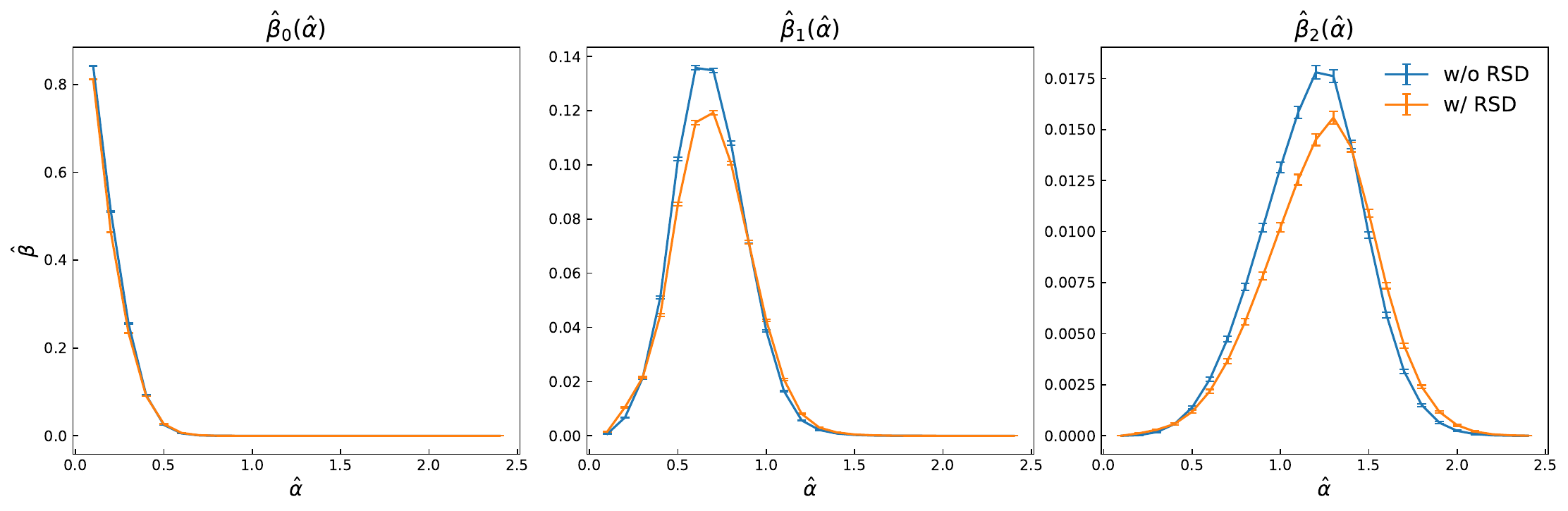}
    \caption{Impact of RSD on Betti curves in fiducial cosmology. The solid lines stand for the average of Betti curves measured from 500 realizations in fiducial cosmology with RSD (orange lines) and without RSD (blue lines). The error bars stand for the standard deviation.
    From the left to the right, they are $\hat{\beta}_0(\hat{\alpha})$, $\hat{\beta}_1(\hat{\alpha})$, and $\hat{\beta}_2(\hat{\alpha})$ respectively.}
    \label{fig:BCrsd}
\end{figure*}

To analyze the impact of RSD on Betti curves, we compare the Betti curves of fiducial simulation with and without RSD in Figure~\ref{fig:BCrsd}. With RSD included, the $\hat{\beta}_0(\hat{\alpha})$ is suppressed. This occurs because $\hat{\beta}_0(\hat{\alpha})$ reflects the number distribution of connected components, and RSD blurs small-scale ($\hat\alpha \lesssim 0.3$) structures. As a result, during filtration, these structures merge earlier into small-scale loops and voids, an effect also manifested as enhancements in $\hat{\beta}_1(\hat{\alpha})$ and $\hat{\beta}_2(\hat{\alpha})$ at the corresponding scales.
At intermediate scales (for $\hat{\beta}_1(\hat{\alpha})$, $0.3 \lesssim \hat\alpha \lesssim 1.0$; for $\hat{\beta}_2(\hat{\alpha})$, $0.3 \lesssim \hat\alpha \lesssim 1.5$), both $\hat{\beta}_1(\hat{\alpha})$ and $\hat{\beta}_2(\hat{\alpha})$ are suppressed, with their peaks reduced in amplitude and shifted toward larger scales
This indicates that the blurring of small-scale connected structures by RSD causes more of them to merge directly into larger connected components during filtration, rather than forming loops or voids. The peak positions of $\hat{\beta}_1(\hat{\alpha})$ and $\hat{\beta}_2(\hat{\alpha})$ correspond to the characteristic scales of dominant loops and voids.
The shift of these peaks toward larger scales (for $\hat{\beta}_1(\hat{\alpha})$, $\hat\alpha \gtrsim 1.0$; for $\hat{\beta}_2(\hat{\alpha})$, $\hat\alpha \gtrsim 1.5$) implies that RSD increases the persistence scale of loops and voids.
Since RSD stretches structures along the line of sight, these features survive filtration to larger scales, which also explains the enhancement of $\hat{\beta}_1(\hat{\alpha})$ and $\hat{\beta}_2(\hat{\alpha})$ at large scales.

\subsection{Sensitivity of Betti curve to cosmological parameters\label{sec:sensitivity}}
\begin{figure*}
    \centering
    \subfigure[Betti curves in different cosmologies without RSD effect.]
    {\label{fig:BCsensitivity_w/o_rsd}
    \includegraphics[width=\linewidth]{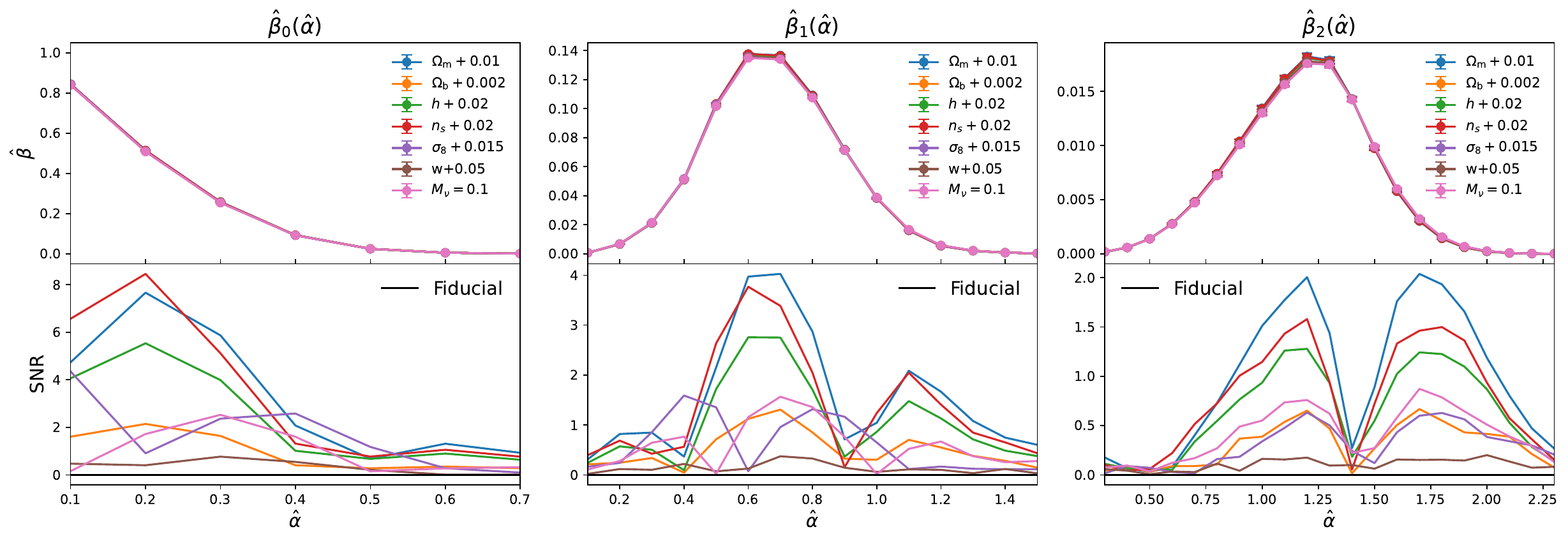}
    }
    \subfigure[Betti curves in different cosmologies with RSD effect.]
    {\label{fig:BCsensitivity_rsd}
    \includegraphics[width=\linewidth]{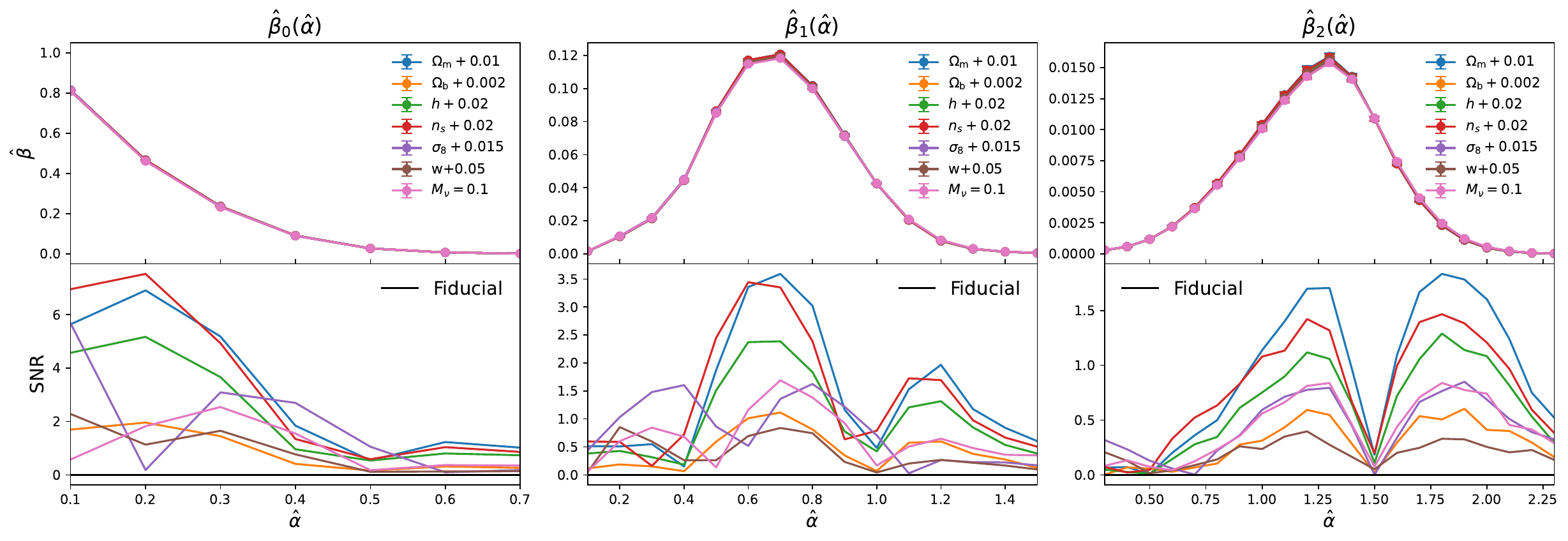}}
    \caption{Betti curves in 0-, 1-, and 2-dimension in several cosmologies (top panel) and their SNR (bottom panel).}
    \label{fig:BCsensitivity}
\end{figure*}

To quantify the dependence of Betti curves on cosmology, we analyze the variations around fiducial sets, where $\Omega_\text{m},\Omega_\text{b}, h, n_{\text{s}},\sigma_{\text{8}},w,\text{and } M_{\nu}$ vary individually. The details of these simulations are listed in Table~\ref{tab:simulation}. We compare Betti curves for different cosmologies in three dimensions and examine their deviations from fiducial cosmology in Figure~\ref{fig:BCsensitivity}. These curves are obtained by averaging Betti curves ($\hat\beta_k$ in dimension $k$) from 500 realizations per cosmology, with errors represented as the standard deviation ($\sigma_{\hat\beta_k}$)\footnote{It is just for visualization. It is the covariance that should be used in scientific analysis.}. For quantitative comparison, we define the signal-to-noise ratio (SNR) of the Betti curve relative to the fiducial cosmology as:
\begin{equation}
    {\rm{SNR}}_{k} = \frac{\langle \Delta\hat\beta_k\rangle}{\sigma_{\Delta\hat\beta_k}},
\label{eq:SNR}
\end{equation}
where 
\begin{equation}
\label{eq:dalta beta}
    \Delta\hat\beta_k =  \hat\beta_k-\hat\beta_k^{\text{fid}},
\end{equation}
is the deviation of the Betti curve for a given cosmology from the fiducial cosmology. Considering the varying initial conditions approximation, we compare the fiducial\_ZA category with simulations whose initial conditions are generated using ZA, whereas for those using 2LPT, we use the fiducial category in Table~\ref{tab:simulation}. 

Figure~\ref{fig:BCsensitivity} shows that the Betti curve can detect percent-level deviations of cosmological parameters near the fiducial cosmology. Given the relatively small volume ($1 \text{Gpc}^3/h^3$), the Betti curves are already sensitive to the variations of cosmology, with the high SNR in certain scale ranges. Cosmological constraints from Betti curve using observational data are expected to be even more promising, given the large survey volume of state-of-the-art spectroscopic surveys.

\begin{figure*}
  \centering
  \subfigure[Parameter derivatives of Betti curves without RSD effect.]
    {\label{fig:derivative_w/o_rsd}
    \includegraphics[width=\linewidth]{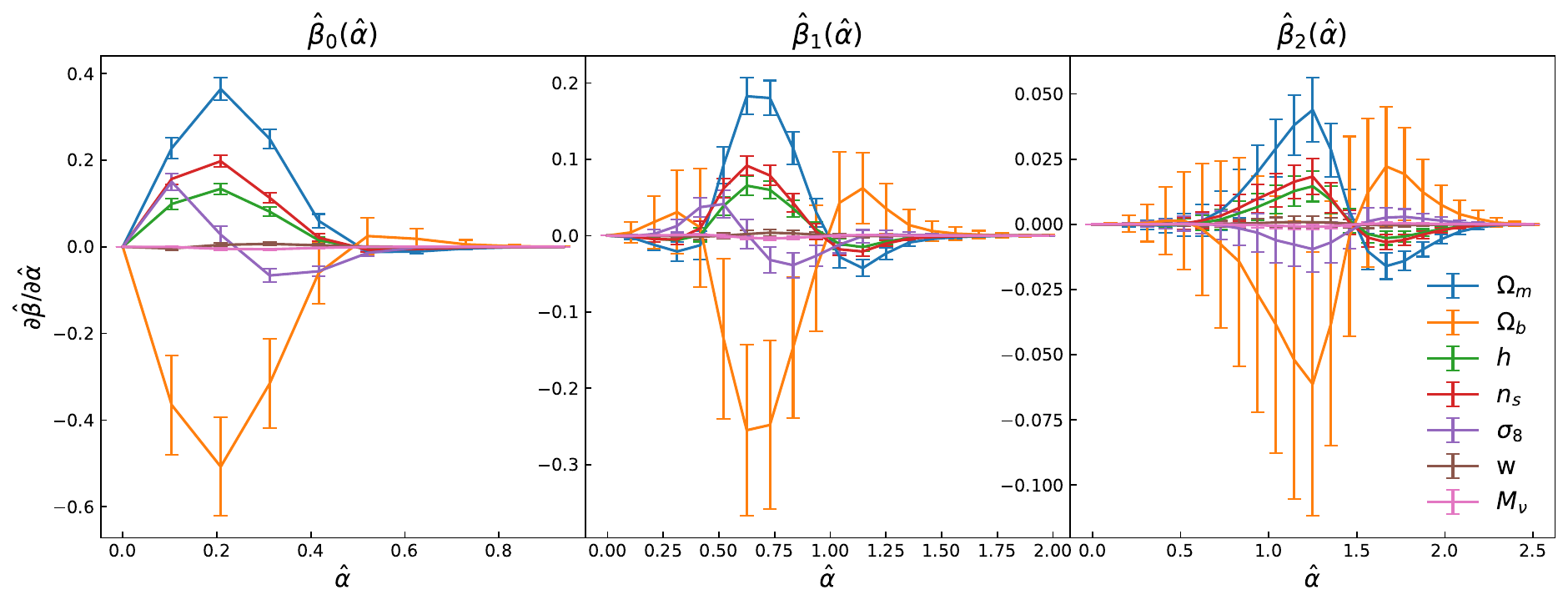}}
  \subfigure[Parameter derivatives of Betti curves with RSD effect.]
    {\label{fig:derivative_w/_rsd}
    \includegraphics[width=\linewidth]{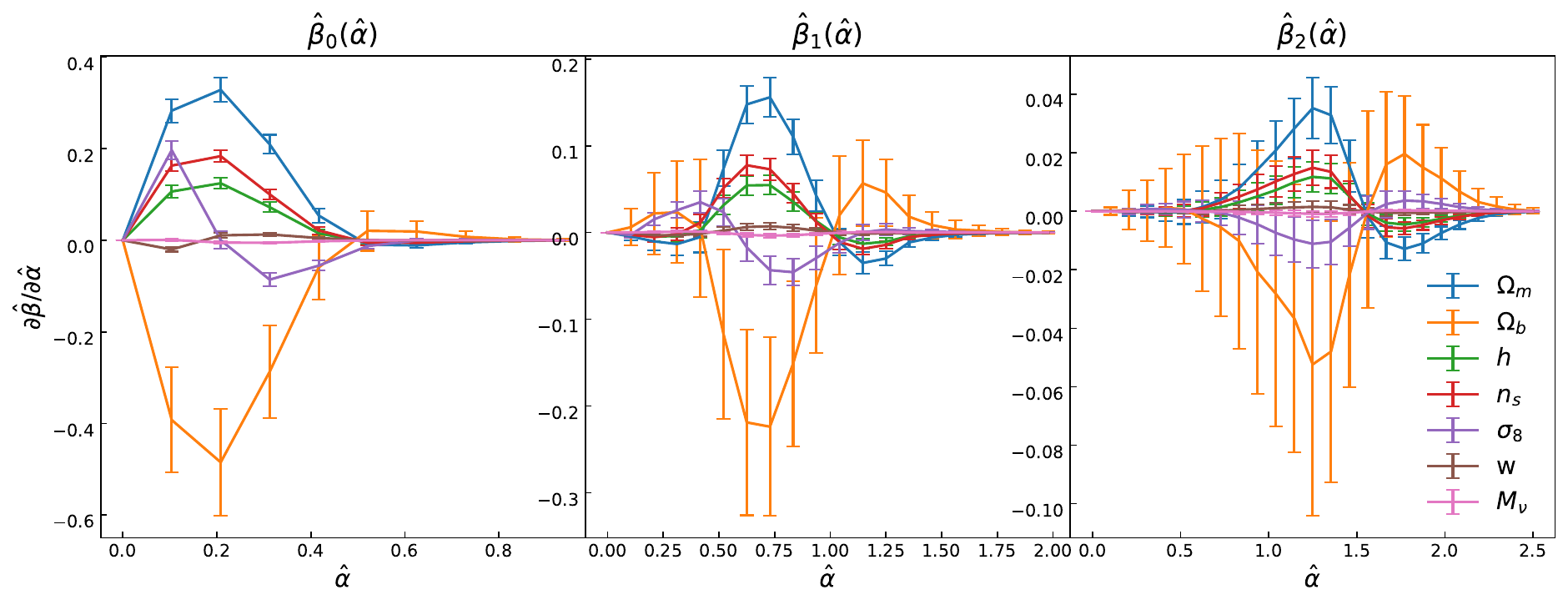}}
  \caption{Parameter derivatives of 0-, 1-, and 2-dimensional Betti curves relative to $ \Omega_\text{m},  \Omega_\text{b}, h,  n_{\text{s}}, \sigma_{\text{8}},w,M_{\nu}$ around fiducial cosmology.\label{fig:derivative}}
\end{figure*}

Since the SNR in Figure~\ref{fig:BCsensitivity} is related to the variation amount of parameters, to further demonstrate that the Betti curve can distinguish different cosmologies, we compute the number derivatives of Betti curves with respect to cosmological parameters $\theta$ around the fiducial cosmology:
\begin{equation}
    \label{eq:derivative}
    \frac{\partial \hat\beta_k}{\partial \theta} = \frac{\hat\beta^{+}_k-\hat\beta^-_k}{2\Delta \theta},
\end{equation}
where $\hat\beta^{\pm}_k$ corresponds to positive or negative variation of a given parameter relative to the fiducial cosmology. For neutrino mass, we use Mnu\_pp for  $\hat\beta^+_k$ and Mnu\_p  for $\hat\beta^-_k$ as listed in Table~\ref{tab:simulation}. The parameter derivatives are visualized in Figure~\ref{fig:derivative}. 

As shown in Figure~\ref{fig:derivative}, Betti curves in all three dimensions are sensitive to the standard cosmological parameters, particularly $\Omega_{\text{m}}$, suggesting their potential to constrain these parameters. This result is broadly consistent with the Fisher forecast of \cite{Cosmo_PH_Fisher}, which also indicates that statistics derived from persistent homology can effectively constrain cosmological parameters. 

It is worth noting that the effects of $\sigma_8$ and $\Omega_{\text{m}}$ on Betti curves act in opposite directions, while the effects of $\Omega_{\text{m}}$ and $n_{\text{s}}$ are similar: increasing $\Omega_{\text{m}}$ or $n_{\text{s}}$ shifts the Betti curve peak to smaller filtration scales with higher amplitudes, whereas increasing $\sigma_8$ delays the peak and lowers its amplitude. This behavior \rev{can be understood qualitatively from} how Betti curves characterize the hierarchical structure of the cosmic web: they trace the abundance and size distribution of structures across scales, reflecting the evolution of large-scale structure. A higher $\Omega_{\text{m}}$ increases the halo number density, producing more and smaller loops and voids that appear earlier and disappear more quickly. The parameter $n_{\text{s}}$, by altering the shape of the primordial power spectrum, redistributes the weight of perturbations across scales; a larger $n_{\text{s}}$ enhances small-scale perturbations, thereby increasing the number of small-scale structures. By contrast, a higher $\sigma_8$ accelerates structure formation, generating larger but sparser loops and voids that persist longer into later stages. Since Betti curves capture the geometric complexity and spatial connectivity of structures, they show high sensitivity to ${\sigma_8, n_{\text{s}}, \Omega_{\text{m}}}$, which directly govern structure formation rates, the distribution of structures across scales, and halo abundance. \rev{It is worth emphasizing that, the interpretation presented here are consistent with large-scale structure phenomenology but are not derived from an analytic topological theory. A fully developed analytic theory for the response of Betti curves to cosmological parameters is not yet available.  To construct the inference pipeline, we have to rely on numerical model of Betti curves, as detailed in the next section.}

\section{Numerical modeling for Betti curve \label{sec:emulator}}

In Section~\ref{sec:sensitivity}, we established that the Betti curves are sensitive to some cosmological parameters. To enable efficient cosmological parameter inference, \rev{we require a forward model that maps cosmological parameters to Betti curves and can be evaluated efficiently within a likelihood framework. However, constructing such a model from first principles is currently infeasible because the filtration process is highly nonlinear and depends on the full geometric and topological evolution of the density field.} From a computational perspective, Bayesian inference for cosmological constraints involves exploring a high-dimensional parameter space and evaluating the likelihood millions of times. Running new simulations at each evaluation step to compute Betti curves is computationally infeasible. \rev{For these reasons, we adopt an emulator approach. The emulator serves as a data-driven surrogate model that interpolates the mapping between cosmological parameters and Betti curves using a finite set of simulations. In the absence of a first-principles analytic model, the emulator is not merely a convenience but a practical necessity for enabling cosmological inference with topological statistics.} In this section, we describe the construction of the data vector and the development of the emulator, \rev{focusing on achieving both predictive accuracy and statistical robustness.}

\subsection{Data vector construction} \label{sec:data vector}
In our pipeline for parameter inference, we estimate the covariance matrix of Betti curves $(C_k)_{ij}$ from the 500 realizations of test cosmology through the sample covariance:
\begin{equation}
    \label{eq:cov}
    (C_k)_{ij} = \frac1{N_{\text{sim}}}\sum_{s=1}^{N_{\text{sim}}}
    [(\hat\beta_k^s)_{i}-\langle\hat\beta_k\rangle_i]
    [(\hat\beta_k^s)_{j}-\langle\hat\beta_k\rangle_j].
\end{equation}
For a $d$-dimensional data vector, the covariance matrix has a size of $d\times d$. To ensure the accuracy of estimation, the number of samples should be significantly larger than the covariance matrix size. Consequently, rather than using the entire Betti curve with 25 data points, which would lead to more covariance components than available samples, we select only the high-SNR regions of the Betti curves as our data vectors, based on Figure~\ref{fig:BCsensitivity} and Figure~\ref{fig:derivative}. 
A basic selection principle is to require the $\rm{SNR} \gtrsim 1$ and maintain the main features of Betti curves.
Specifically, for $\hat\beta_0$, we choose $\hat\alpha \in [0.1, 0.5]$ with a resolution of $\Delta\hat\alpha=0.1$. The first point at $\hat\alpha=0$ is omitted since $\hat\beta_0$ is exactly one there, carrying no cosmological information and only statistical noise. For $\hat\beta_1$, the cosmological information is mainly carried by the amplitude of peak and the slope of the larger scale region. Thus, we select $\hat\alpha \in [0.2,1.4]$, which includes the two highest SNR ranges of the Betti curve containing most cosmological information. For $\hat\beta_2$, in addition to peak amplitude and slope of region over peak scale, the slope of region below peak scale also carries information about voids merging. However, the small-scale features of $\hat\beta_2$ may be influenced by non-cosmological effects, such as the nonlinear halo formation process, which is beyond our scope. To balance cosmological information and on-cosmological effects, we choose $\hat\alpha \in [0.9, 1.8]$.
\rev{We emphasize that this data vector definition is not optimized to extract maximal information, but rather chosen to demonstrate the feasibility of Betti-curve–based inference under conservative conditions. A fully information optimal data compression scheme is left for future work.}

\subsection{Automated machine learning emulator\label{sec:automl}}
Automated Machine Learning (AutoML) aims to significantly lower the expertise barrier and tuning cost of building machine learning models by automating model selection, feature engineering, and hyperparameter optimization.
We employ the AutoML framework \texttt{auto-sklearn} \cite{auto-sklearn}, built upon the \texttt{Python} machine learning library \texttt{scikit-learn} \cite{scikit-learn}, to construct emulators for Betti curves in each dimension, both with and without the effect of RSD. The core workflow of \texttt{auto-sklearn} can be summarized as follows. It defines a search space that combines various preprocessing methods, such as Principal Component Analysis (PCA \cite{PCA}), Independent Component Analysis (ICA \cite{ICA}), and Polynomial Feature expansion (PF \cite{scikit-learn}), with multiple regressors, including Random Forests \cite{RandomForest}, Gradient Boosted Regression Trees (GBRT \cite{GDBT}), and Gaussian Process Regression (GPR \cite{GPR}), together with their associated hyperparameters. Leveraging meta-learning \cite{meta-learning} for warm-start initialization, it performs iterative pipeline evaluation using Bayesian optimization based on Sequential Model-based Algorithm Configuration (SMAC \cite{SMAC}), gradually refining candidate pipelines according to their validation performance. The outcome is either a single optimized pipeline or an ensemble of top-performing pipelines combined into the final predictive model.

To construct the emulators for Betti curves, we shuffle the 2000 cosmologies from the nwLH simulations and randomly split them into a training set (1800 cosmologies) and a test set (200 cosmologies). The training set is used for model fitting, while the test set evaluates predictive accuracy and guides model optimization. The input data vector $\boldsymbol{\theta}$ includes both cosmological parameters and a nuisance parameter $\ell$, defined as the mean halo separation in a given catalog (see Section~\ref{sec:normalize}). This parameter characterizes the halo number density within a fixed volume and influences the overall amplitude of Betti curves. Introducing $\ell$ helps suppress non-cosmological systematic effects, thereby improving emulator performance.

During training, we first perform Bayesian optimization with \texttt{auto-sklearn} without pre-specifying the model type, searching across different preprocessing methods, feature selection techniques, and regression algorithms. The results show that, across cross-validation, GPR consistently provides the best performance for Betti curves of order 0, 1, and 2. This reflects the inherent advantage of GPR in medium-scale ($\sim 10^3$), low-dimensional ($\lesssim 10$) regression tasks where the target functions are smooth or exhibit well-defined local peak–valley structures. By leveraging kernel functions, GPR globally models both correlations and uncertainties in the data (see \citet{GPR} for details). Building on this result, we fix GPR as the regression model and reapply \texttt{auto-sklearn} to search for the optimal preprocessing and feature engineering methods tailored to different orders of Betti curves. \rev{The preprocessing strategies selected by \texttt{auto-sklearn} reflect statistical properties of the measured curves and can be empirically explained as follows:
\begin{itemize}
\item For $\hat\beta_0$, Independent Component Analysis (ICA)\footnote{ICA decomposes multivariate observations into statistically independent non-Gaussian components. Unlike PCA, which identifies orthogonal directions with maximum variance, ICA maximizes non-Gaussianity (e.g., kurtosis, entropy) to recover independent sources. Formally, ICA assumes the observed data $\mathbf{X}$ are linear mixtures of independent sources $\mathbf{S}$ via an unknown mixing matrix $\mathbf{A}$, i.e., $\mathbf{X} = \mathbf{A}\mathbf{S}$. The goal is to estimate an unmixing matrix $\mathbf{W}\approx \mathbf{A}^{-1}$ such that $\mathbf{S} \approx \mathbf{W}\mathbf{X}$. See \citet{ICA} for details.} is favored. Empirically, $\hat\beta_0$ exhibits strong correlations between adjacent filtration samples and a relatively smooth, monotonic trend over the selected interval. ICA separates redundant signals into statistically independent components, achieving denoising and dimensionality reduction, thus providing cleaner inputs for GPR.
\item For $\hat\beta_1$ and $\hat\beta_2$, Polynomial Feature expansion (PF)\footnote{PF expands the original input variables into polynomial combinations (e.g., quadratic, cross terms), enriching feature representation to capture nonlinear interactions. This effectively projects the data into a higher-dimensional space, enabling the model to learn nonlinear relationships. See \citet{scikit-learn} for implementation.} is selected. These curves display pronounced nonlinear peak–valley structures in $\hat\alpha$-space. PF enriches the feature set by generating higher-order polynomial terms and interactions, allowing GPR to accurately capture both local extrema and global trends in the curves.
\end{itemize}
}

In summary, \texttt{auto-sklearn} not only automatically identifies the optimal regression model for all Betti curves but also selects preprocessing strategies that align with their statistical characteristic. This data-driven modeling process improves emulator accuracy, provides insights into structural differences across Betti curve orders, and offers guidance for future physics-informed modeling and feature engineering strategies.

\subsection{Emulator performance \label{sec:emulator performance}}
\begin{figure*}
    \subfigure[Simulated and Emulated Betti curves not including RSD]{
    \includegraphics[width=\linewidth]{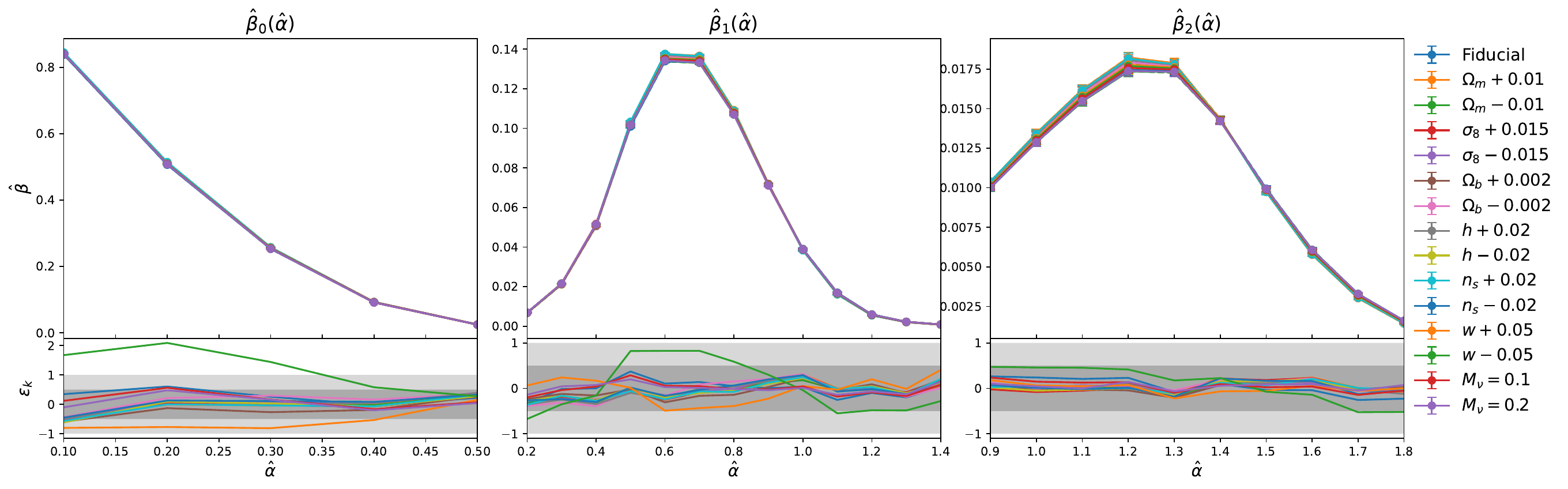}
    \label{fig:BCvalidation_w/o_rsd}
    }
    \subfigure[Simulated and Emulated Betti curves including RSD]{
        \includegraphics[width=\linewidth]{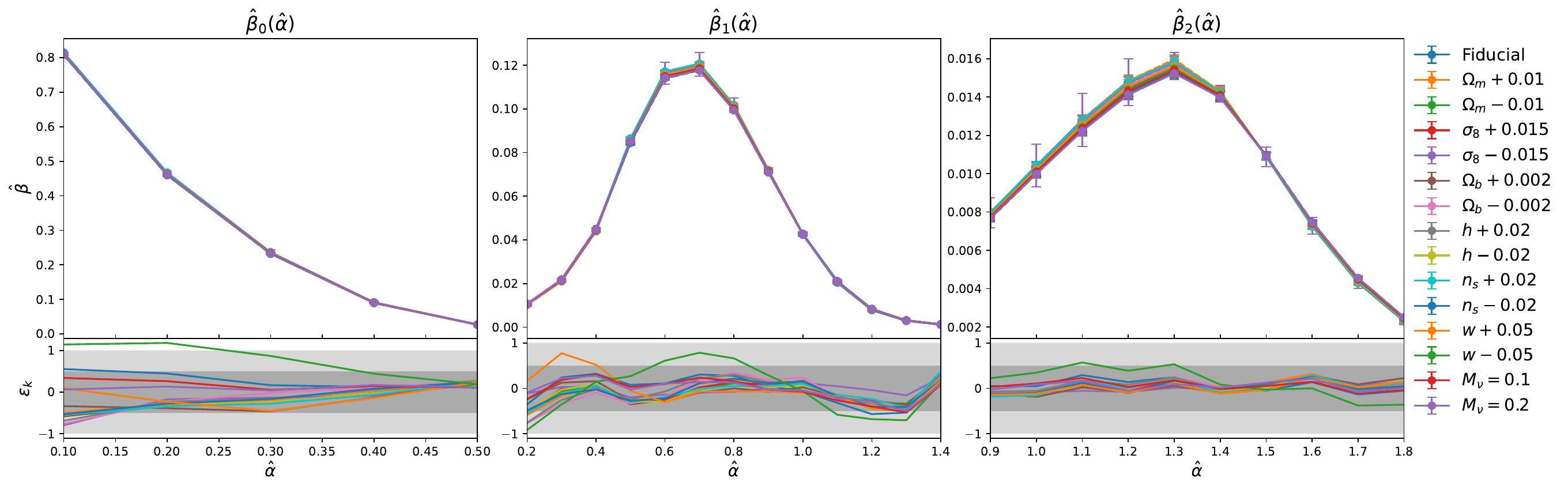}
    \label{fig:BCvalidation_w/_rsd}
    }
\caption{Performance of emulators for 0-, 1-, and 2-dimensional Betti curves for serval test cosmologies. Top panel: The solid lines with error bars stand for the Betti curves measured from the simulation, and the dashed lines stand for the prediction of emulators.  Bottom panel: The prediction errors of emulators for test cosmologies. The gray band represents the region where the prediction errors are within 0.5 (darker) and 1 (lighter).\label{fig:BC_val}}
\end{figure*}

\begin{figure}
    \includegraphics[width=\linewidth]{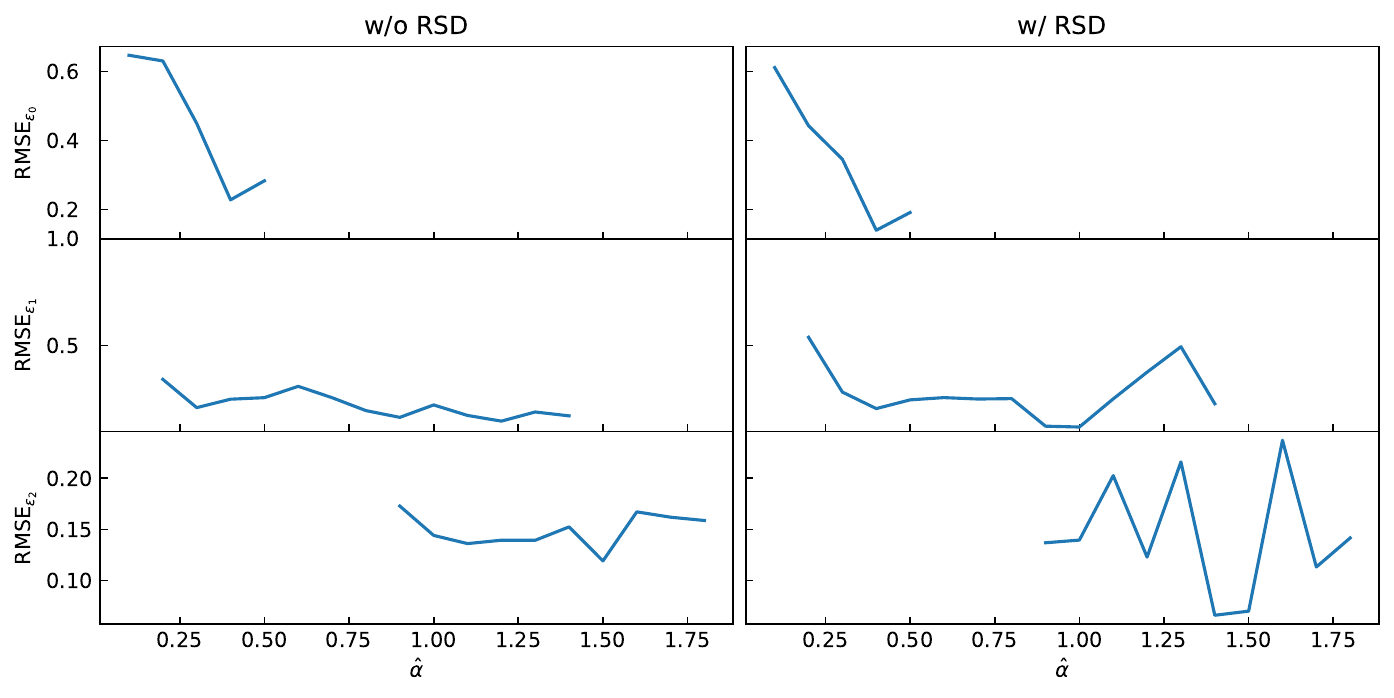}
    \caption{The RMSE for 0-, 1-, and 2-dimensional Betti curve emulators from top to bottom. The left plot does not include RSD, while the plot on the right does.}
    \label{fig:rmse}
\end{figure}

The validation cosmologies used in our work include the fiducial cosmology and seven individually varying parameter sets corresponding to $\{ \Omega_\text{m},  \Omega_\text{b}, h,  n_{\text{s}}, \sigma_{\text{8}},w,M_{\nu}\}$, resulting in 13 validation sets in total. Each cosmology has 500 realizations. For each realization, we predict Betti curves using the trained emulators. The final emulator prediction for a given cosmology is obtained by averaging over the 500 individual predictions. Similarly, the measured Betti curves are computed as the mean Betti curve from 500 realizations, with the standard error estimated from the same set.
To quantify the accuracy of the emulator, we define the relative prediction error $\epsilon_k$ for each validation set as:
\begin{equation}
    \label{eq:err}
    \epsilon_k(\hat\alpha;\boldsymbol{\theta}) = \frac{\langle\hat\beta^{\text{pred}}_k\rangle_{\boldsymbol{\theta}}-\langle\hat\beta^{\text{obs}}_k\rangle_{\boldsymbol{\theta}}}{\sigma_{\hat\beta_k}},
\end{equation}
where $\langle\hat\beta^{\text{pred}}_k\rangle_{\boldsymbol{\theta}}$ is the emulator's prediction, $\langle\hat\beta^{\text{obs}}_k\rangle_{\boldsymbol{\theta}}$ is the measured Betti curve, and $\sigma_{\hat\beta_k}$ is the statistical error.
The result, shown in Figure~\ref{fig:BC_val}, indicates that for all validation cosmologies, the emulator predictions remain within approximately 0.5 $\sigma_{\text{stat}}$ of the measured values. 
The intrinsic fluctuation of training cosmology from the single realization and small volume, which is close to $\sigma_{\text{stat}}$, limits the performance of the emulator. \rev{The uncertainty of emulator may also introduce bias into cosmological constraints. To account for this effect, we propagate emulator uncertainty into parameter inference by incorporating the emulator error into the total covariance matrix used in the likelihood analysis.}
\rev{Nevertheless, these limitations are expected to be mitigated by training on simulations with larger volumes and multiple realizations per cosmology.
As a proof-of-concept, the current emulator's accuracy is sufficient for demonstrating Betti curves' potential in constraining cosmology.}
To further assess emulator performance, we compute the Root Mean Square Error (RMSE) of $\epsilon_k$ ($\text{RMSE}_{\epsilon_k}$) across all validation cosmologies:
\begin{equation}
    \label{eq:RMSE}
    \text{RMSE}_{\epsilon_k}(\hat\alpha) = \sqrt{\frac1{N_{\text{val}}}\sum_{i=1}^{N_{\text{val}}}\epsilon_k^2(\hat\alpha;\boldsymbol{\theta}_i)},
\end{equation}
where $N_{\text{val}}$ is the number of validation cosmologies. 
The results in Figure~\ref{fig:rmse} confirm that for all three emulators, $\text{RMSE}_{\epsilon_k}<1$ across all $\hat\alpha$ scales, indicating robust performance near the fiducial cosmology.

\section{Results \label{sec:result}}
\subsection{Bayesian inference framework \label{sec:Bayesian}}
We perform Bayesian inference for parameter recovery using nested sampling to sample the posterior distribution. We assume flat priors with the same parameter ranges as the training set for cosmological parameters and $[0,100]$ for the nuisance parameter $\ell$, which is sufficiently large given that the typical $\ell$ for our halo catalogs is about $10$. 
\rev{We employ a Gaussian likelihood for our analysis. However, because Betti curves are derived from counting statistics, we also considered a Poisson likelihood. The resulting posterior showed no statistically significant difference due to the large number counts, but the computational cost of Poisson likelihood was nearly ten times higher. Therefore, we adopted the Gaussian likelihood as follows for its efficiency:}
\rev{\begin{equation}
    \label{eq:likelihood}
    \text{log}\mathcal{L}(\boldsymbol{d}|\boldsymbol{\theta}) = -\frac1{2}\cdot\frac{n-p-2}{n-1}\cdot
    [\boldsymbol{d}-\boldsymbol{m}(\boldsymbol{\theta})]^TC^{-1}_{\rm tot}[\boldsymbol{d}-\boldsymbol{m}(\boldsymbol{\theta})],
\end{equation}
where $\boldsymbol{d}$ is data vector, $\boldsymbol{m}(\boldsymbol{\theta})$ is emulator's prediction, and $C_{\rm tot}$ is the total covariance matrix.}
The prefactor $\frac{n-p-2}{n-1}$ accounts for the unbiased inverse covariance matrix estimator as suggested by \cite{cov}, where $n$ is the total number of realizations and $p$ is the dimension of the data vector.
\rev{The total covariance includes both statistical fluctuations of the Betti curves and systematic uncertainty associated with emulator predictions:
\begin{equation}C_{\rm tot} = C_{\text{stat}}+C_{\text{emu}}.\end{equation}
To estimate the emulator-induced uncertainty, we use the test set of 200 cosmologies. For each cosmology in the test set, we compute the residual vector
\begin{equation}\boldsymbol{r} = \boldsymbol{d}_{\rm test}-\boldsymbol{m}_{\rm test}.\end{equation}
Assuming that the covariance of Betti curves and the emulator prediction error is independent of cosmology (the assumption is validated in Section~\ref{sec:fid result}) the covariance of these residuals $C(\boldsymbol{r})$ can be approximately written as 
\begin{equation}\label{eq:Cov(r)}C(\boldsymbol{r})\approx C_{\text{fid}}+C_{\text{emu}}\approx C_{\rm tot},\end{equation}
where $C_{\text{fid}}$ is the covariance estimated from the 500 realizations of the fiducial cosmology (Equation~\ref{eq:cov}).
We therefore obtain an estimate of the emulator covariance as
\begin{equation}C_{\rm emu} \approx \rm diag \left(C(\boldsymbol{r})-C_{\rm fid}\right),\end{equation}
where $\rm diag()$ represent diagonal matrix. Because no clear off-diagonal residuals are observed, though there are moderate statistical fluctuations due to the limited test set size, we include only the diagonal contribution from $C_{\rm emu}$ and retain the off-diagonal structure from $C_{\rm fid}$. The final covariance used in the likelihood analysis is
\begin{equation}C_{\rm tot} = C_{\rm fid} + \rm diag \left(C(\boldsymbol{r})-C_{\rm fid}\right).\end{equation}
}
We derive posterior probability distributions with the nested sampling Monte Carlo algorithm MLFriends \cite{MLFriends14,MLFriends19} using the
\texttt{UltraNest}\footnote{\url{https://johannesbuchner.github.io/UltraNest/}} package \cite{UltraNest}. The nested sampling algorithm explores parameter space globally in an unsupervised manner, proceeding without problem-specific tuning until reaching a well-defined convergence \cite{NestedSampling23}. Compared to Markov Chain Monte Carlo (MCMC), nested sampling is better suited for complex, high-dimensional posteriors with nonlinear parameter correlations.

We compare the cosmological constraints from Betti curves with and without RSD effect, and also present the constraints from combining Betti curves and power spectrum.
Beyond the fiducial box size (1 Gpc/$h$), we also conduct parameter recovery tests for Betti curves on sub-box simulations to assess the influence of cosmic variance on inference results. 

\subsection{Fiducial result \label{sec:fid result}}
We evaluate the constraining power of $\hat\beta_0$, $\hat\beta_1$, $\hat\beta_2$, and their combination on the fiducial cosmology, both with and without RSD. We further conduct a joint analysis of Betti curves and the power spectrum, and compare the results obtained with and without the inclusion of RSD.
Betti curves and power spectrum are computed from 500 realizations, with the mean curves adopted as the observational data. Trained emulators are employed as the theoretical models for Betti curves. For the power spectrum, we similarly construct emulators following the methodology outlined in Section~\ref{sec:emulator}. The likelihood is then evaluated using Equation~\ref{eq:likelihood}.
\subsubsection{Constraints from Betti curves\label{sec:fid_BC}}
\begin{figure*}
    \includegraphics[width=\linewidth]{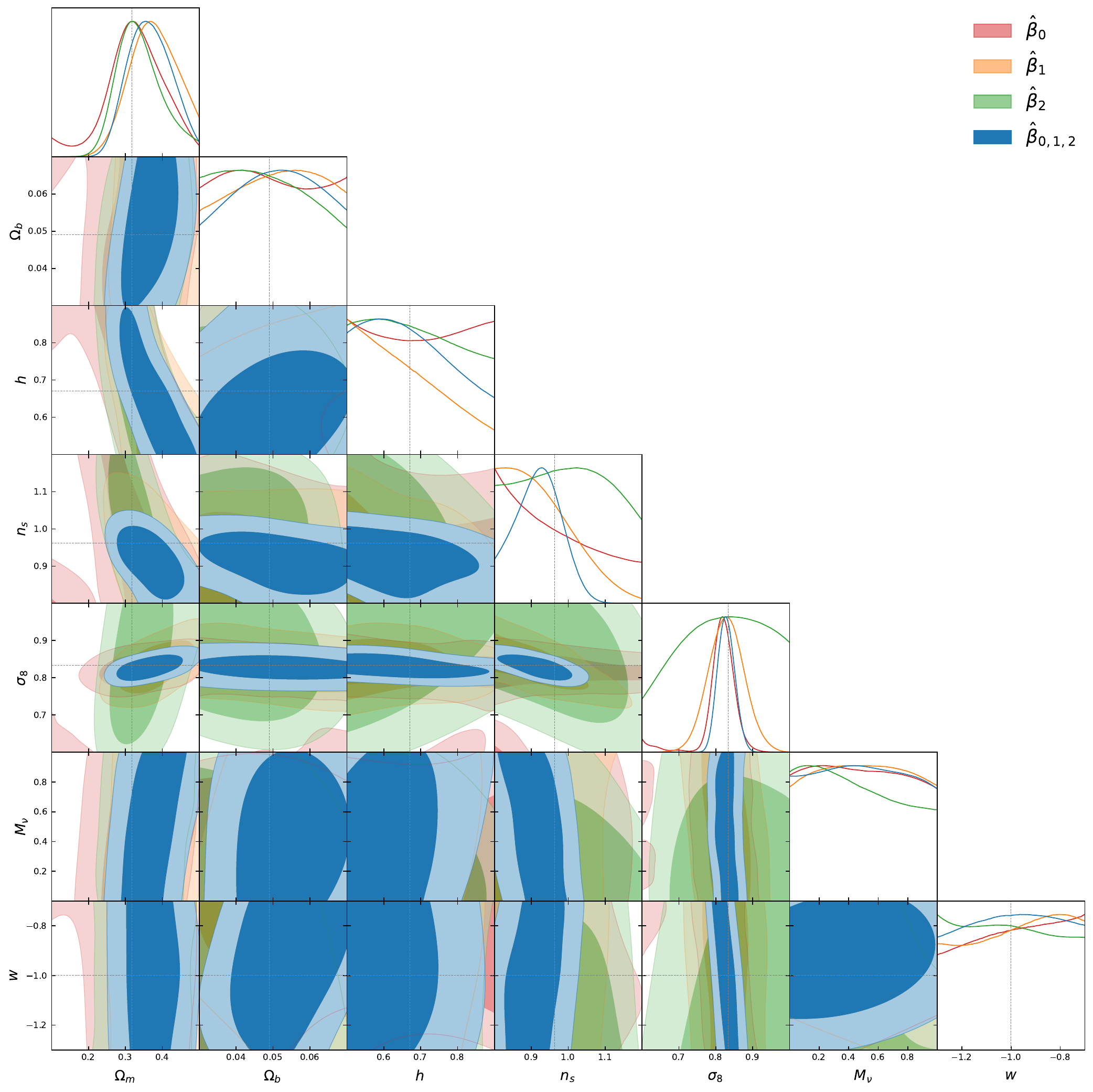}
    \caption{The posterior distribution for cosmological parameters  $ \Omega_\text{m},  \Omega_\text{b}, h,  n_{\text{s}}, \sigma_{\text{8}}, w, M_{\nu}$ in fiducial cosmology with fiducial box size. The contours stand for the recovered posterior from 0- (red), 1- (orange), 2-dimensional (green) Betti curves, and their combination (blue). The contours mark 68\% (1-$\sigma$) and 95\% (2-$\sigma$) regions of the posteriors. The crossed dashed lines mark the true values for the parameters.}
    \label{fig:BCs_fza}
\end{figure*}
In the absence of RSD effects, the posterior distributions of cosmological parameters constrained by $\hat\beta_0$, $\hat\beta_1$, $\hat\beta_2$, and their combination under the fiducial cosmology are shown in Figure~\ref{fig:BCs_fza}. The results indicate that Betti curves provide the strongest constraints on ${\sigma_8, n_{\text{s}}, \Omega_{\text{m}}}$, followed by ${h,\Omega_{\text{b}}}$, while their constraining power on $w$ is weaker, and essentially negligible for $M_{\nu}$. This conclusion is consistent with the sensitivity analysis in Section~\ref{sec:sensitivity}, as well as with the findings of \citet{PDCNN24}.

\rev{Considering the parameters ${\sigma_8, n_{\text{s}}, \text{and } \Omega_{\text{m}}}$, among the individual Betti curves, $\hat\beta_1$ delivers the strongest constraints overall. $\hat\beta_0$ and $\hat\beta_2$ provide weaker constraints in most cases, with the notable exception of $\sigma_8$, for which $\hat\beta_0$ yields the tightest constraint while $\hat\beta_2$ carries almost no constraining power.} This reflects the balance between signal-to-noise ratio and parameter sensitivity. $\hat\beta_0$ tracks the number of connected components and is highly responsive to density fluctuations governed by $\sigma_{\text{8}}$, and its large sample count naturally reduces statistical uncertainty. However, it contains relatively limited cosmological information. By contrast, $\hat\beta_2$ captures void information, carrying richer information but suffering from higher noise due to the scarcity of voids, leading to weaker constraining power than $\hat\beta_1$. 
\rev{These findings are broadly consistent with the recent analysis, which reported that most of the cosmological information in Betti curves is contained in $\beta_0$ and $\beta_1$, with $\beta_2$ contributing little standalone constraining power \cite{BC012power}. In our results, $\hat\beta_2$ indeed provides weak constraints when considered individually. However, it contributes non-negligibly when combined with the other Betti curves by different parameter degeneracy directions, thereby modestly tightening joint constraints. This suggests that while $\beta_2$ is not a dominant information carrier on its own, it can still play a complementary role in multi-statistic analyses.}

Overall, the posterior scatter of the nuisance parameter $\ell$ is substantially smaller than that of other parameters, while all cosmological parameters except $M_{\nu}$ are recovered without bias, confirming that the emulator successfully extracts cosmological information encoded in Betti curves. 
For the neutrino mass $M_{\nu}$, however, Betti curves show weak sensitivity. \rev{On the one hand, Massive neutrinos primarily influence small-scale structures, while Betti curves mainly encode the connectivity and topology of collapsed structures across intermediate scales. At the simulation volume and resolution considered here, this topological response appears subdominant compared to statistical fluctuations and emulator uncertainties. On the other hand,} combined with the physical requirement of non-negative mass, and the fact that the training set only includes cosmologies with $M_{\nu} \geq 0$, the emulator cannot fully learn the influence of $M_{\nu}$ on Betti curves, ultimately preventing it from providing tight constraints.

\rev{
To test the assumption made in Equation~\ref{eq:Cov(r)} that the covariance of Betti curves is independent of cosmology, we perform an explicit validation. We treat the fiducial set as the mock observation and repeat the parameter recovery using covariance matrices estimated separately from the fiducial set, the s8\_m set, and the s8\_p set.}
\rev{The s8\_m and s8\_p cosmologies differ from the fiducial value of $\sigma_8$ by approximately 3.6\%, while the marginalized 1-$\sigma$ uncertainty on $\sigma_8$ from Betti curves is about 2.9\%. }

\begin{figure*}
    \includegraphics[width=\linewidth]{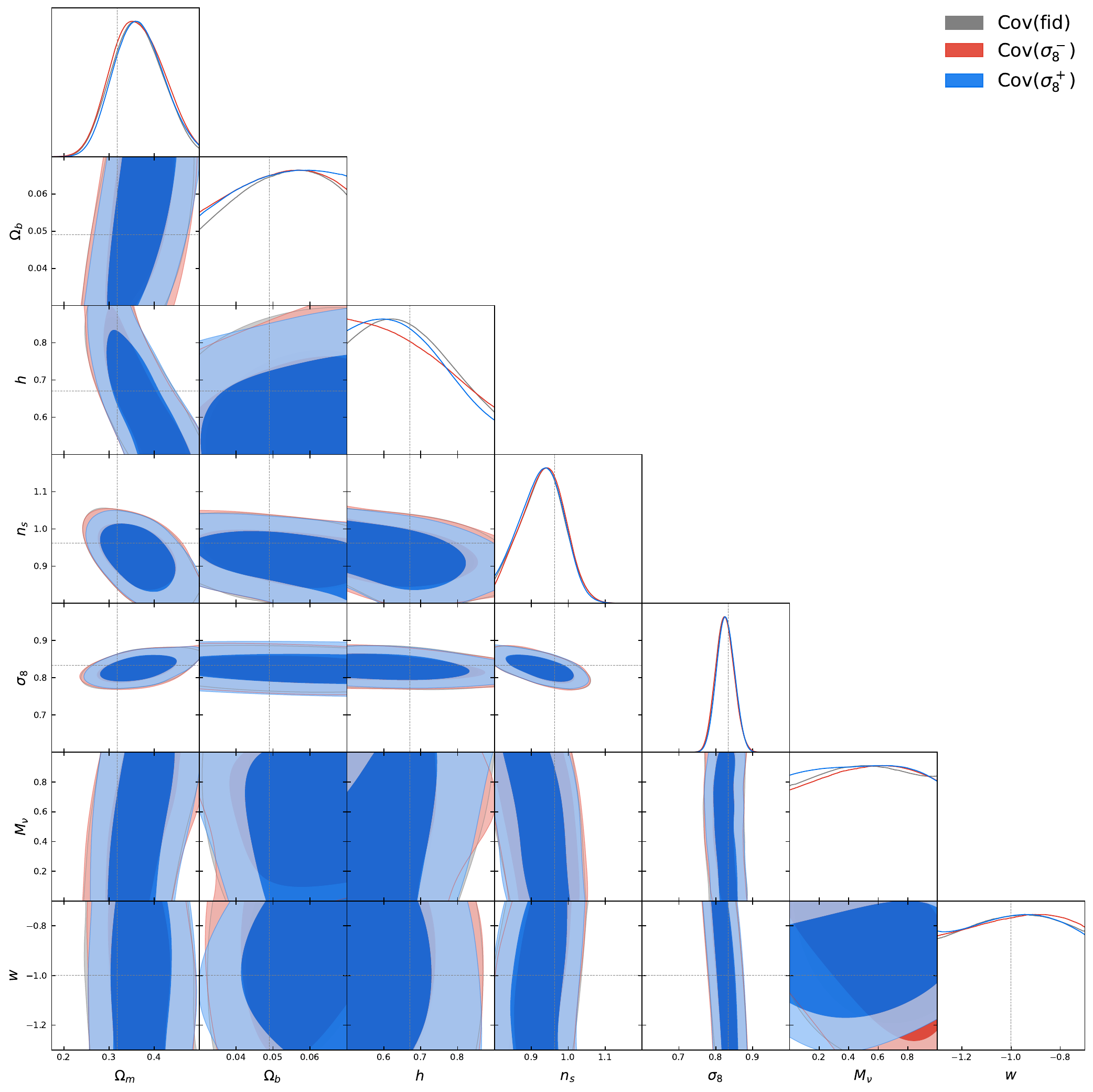}
    \caption{Parameter constraints from the
combination of Betti curves $\hat\beta_0$, $\hat\beta_1$, $\hat\beta_2$ under fiducial cosmology with covariance estimated from fiducial set (gray contours), s8\_m set (red contours), and s8\_p set (blue contours).}
\label{fig:BCs_fza_compare_cov}
\end{figure*}

\rev{Figure~\ref{fig:BCs_fza_compare_cov} shows that the recovered parameter constraints are highly consistent across all three covariance choices. The posterior contours overlap within statistical fluctuations, with no detectable bias or systematic shift. The small difference indicates that, at the current precision level and for the explored parameter range, the cosmology dependence of the Betti-curve covariance does not significantly affect parameter inference.
}

\begin{figure*}
    \includegraphics[width=\linewidth]{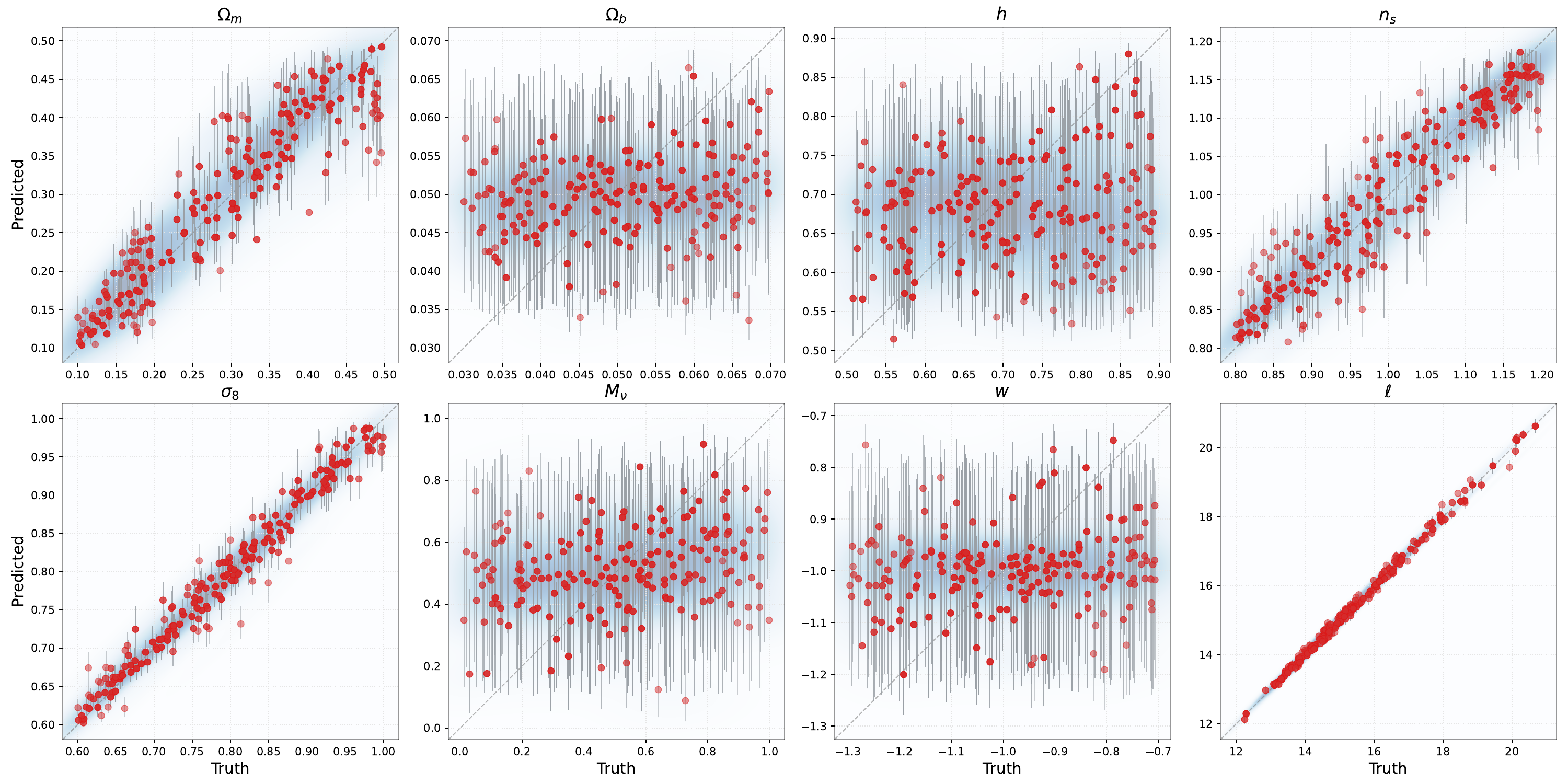}
    \caption{Recovered versus true values for cosmological and nuisance parameters of test set without the inclusion of RSD. Red points are the measurement of test cosmologies. The gray error bar marks the 1-$\sigma$ region of one measurement. The blue background highlights the distribution of (Truth, Recovered), calculated through a 2D kernel density estimation. The dashed diagonal indicates the one-to-one relation.}
    \label{fig:nwLH_recovery_fof}
\end{figure*}

To assess the statistical stability of the inference pipeline, we perform parameter recovery experiments for each of the 200 realizations in the test set, as shown in Figure~\ref{fig:nwLH_recovery_fof}. For ${\Omega_{\text{m}}, \sigma_8, n_{\text{s}}}$, the recovered values exhibit a tight clustering along the one-to-one (Truth, Recovered) line, indicating that Betti curves provide strong constraining power on these parameters, which is consistent with the results presented in Figure~\ref{fig:BCs_fza}. This consistency further supports the reliability of the inference pipeline.

\subsubsection{The effect of RSD to constraints from Betti curves \label{sec:fid_BC_RSD}}
\begin{figure*}
    \includegraphics[width=\linewidth]{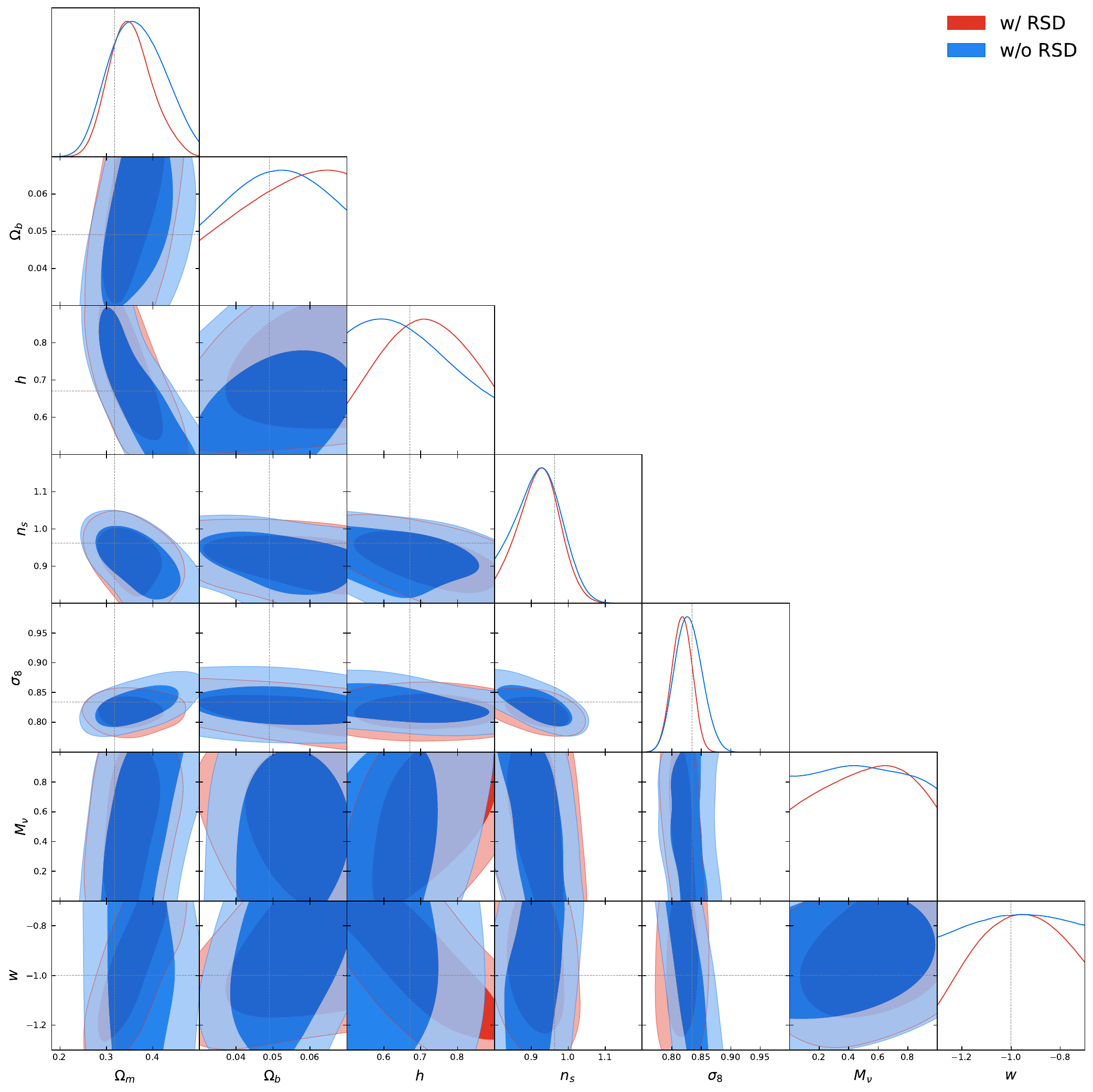}
    \caption{Parameter constraints under fiducial cosmology with (red contours) and without (blue contours) RSD from the combination of Betti curves $\hat\beta_0$, $\hat\beta_1$, $\hat\beta_2$.
    The contours mark 68\% (1-$\sigma$) and 95\% (2-$\sigma$) regions of the posteriors. The crossed dashed lines mark the true values for the parameters.}
    \label{fig:BCs_fza_rsd}
\end{figure*}
Figure~\ref{fig:BCs_fza_rsd} presents the joint cosmological parameter constraints from Betti curves under fiducial cosmology, with and without RSD.
Except for $M_{\nu}$, which Betti curves cannot effectively constrain, the inclusion of RSD still yields unbiased parameter estimates.
Among all constrained parameters, the most significant improvement appears in $\sigma_8$, for which the constraining power increases by \rev{27\%.}
This enhancement arises because $\sigma_8$ determines the amplitude of the linear matter power spectrum, and the RSD effect is directly governed by the overall fluctuation amplitude.
Consequently, incorporating RSD information greatly strengthens the sensitivity of Betti curves to $\sigma_8$, leading to substantially tighter constraints \cite{RSD2S8}. The inclusion of RSD also improves constraints on $\Omega_{\text{m}}$ and $w$ by about \rev{19\% and 10\%,} respectively. For $\Omega_{\text{m}}$, RSD originates from both large-scale coherent flows and small-scale random motions, both of which are closely linked to the matter density.
The small-scale Fingers-of-God (FoG) effect reflects the velocity dispersion within clusters, while the large-scale Kaiser effect \cite{KaiserEffect} traces the global growth rate $f \propto \Omega_{\text{m}}^{0.55}$ \cite{growthfactor}. Together, these components provide additional constraints on $\Omega_{\text{m}}$.
For $w$, which governs the late-time cosmic acceleration and structure growth rate, the RSD effect carries information about late-time growth, thereby enhancing the constraints on $w$.
In contrast, for $h$, $\Omega_{\text{b}}$, and $n_{\text{s}}$, the inclusion of RSD yields negligible improvement. This is expected because $h$ primarily affects the background expansion and overall spatial scale, $\Omega_{\text{b}}$ influences the small-scale baryonic composition, and $n_{\text{s}}$ determines the shape of the primordial power spectrum, all of which are weakly correlated with the velocity-field information and spatial anisotropies introduced by RSD.

\begin{figure*}
    \includegraphics[width=\linewidth]{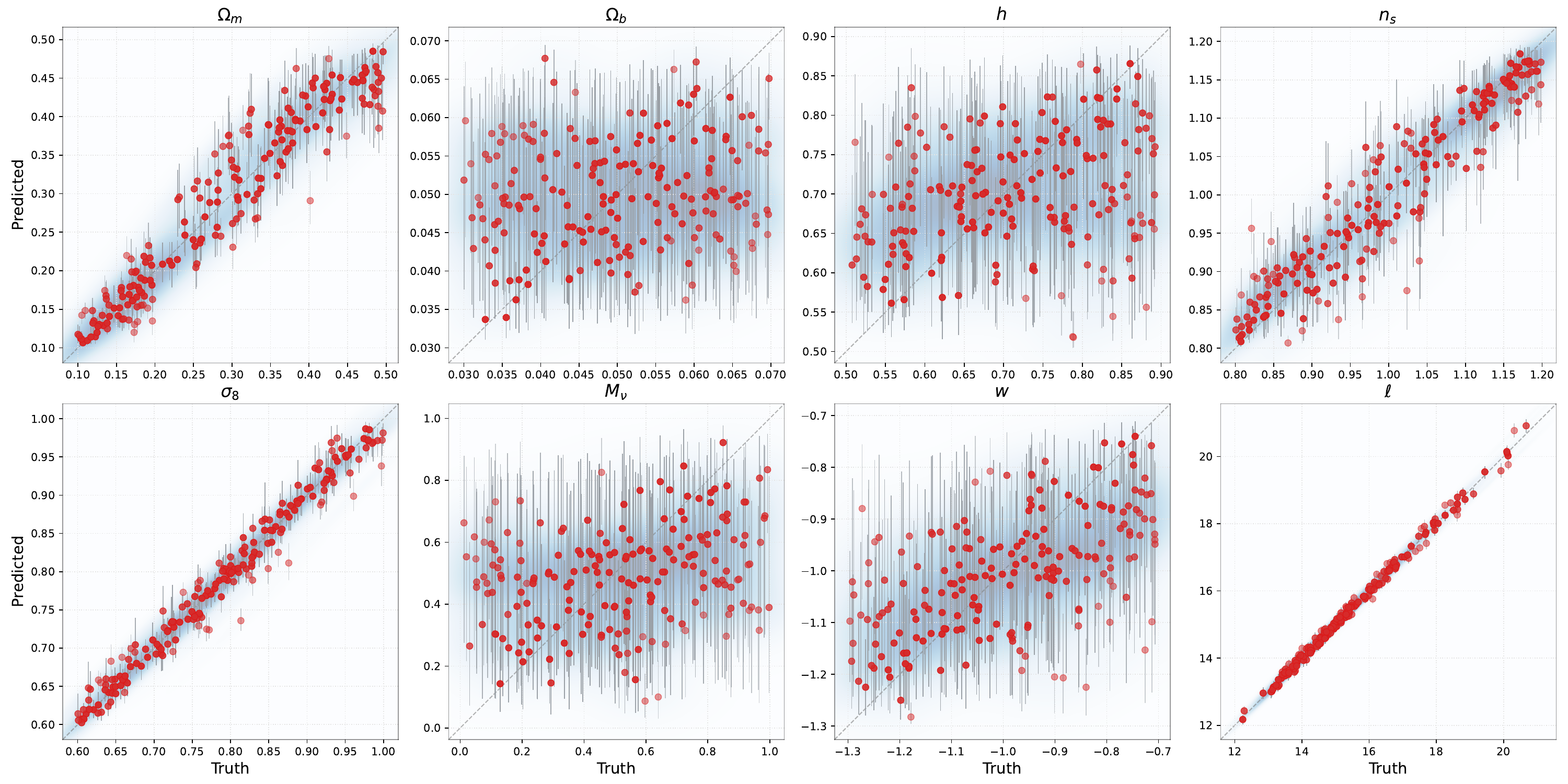}
    \caption{Recovered versus true values for cosmological and nuisance parameters of test set with the inclusion of RSD. Red points are the measurement of test cosmologies. The gray error bar marks the 1-$\sigma$ region of one measurement. The blue background highlights the distribution of (Truth, Recovered), calculated through a 2D kernel density estimation. The dashed diagonal indicates the one-to-one relation.}
    \label{fig:nwLH_recovery_fof_rsdz}
\end{figure*}

We also perform the same parameter recovery experiment in the presence of RSD, as shown in Figure~\ref{fig:nwLH_recovery_fof_rsdz}. The parameters ${\Omega_{\text{m}}, \sigma_8, n_{\text{s}}}$ remain well constrained, while the (Truth, Recovered) distribution of $w$ becomes more tightly clustered along the diagonal. This trend indicates that incorporating RSD enhances the sensitivity of Betti curves to $w$, and the consistent performance confirms the reliability of the emulators in redshift space.

\subsubsection{Joint analysis of Betti curves and power spectrum \label{sec:fid_BC_Pk}}
\begin{figure*}
    \includegraphics[width=\linewidth]{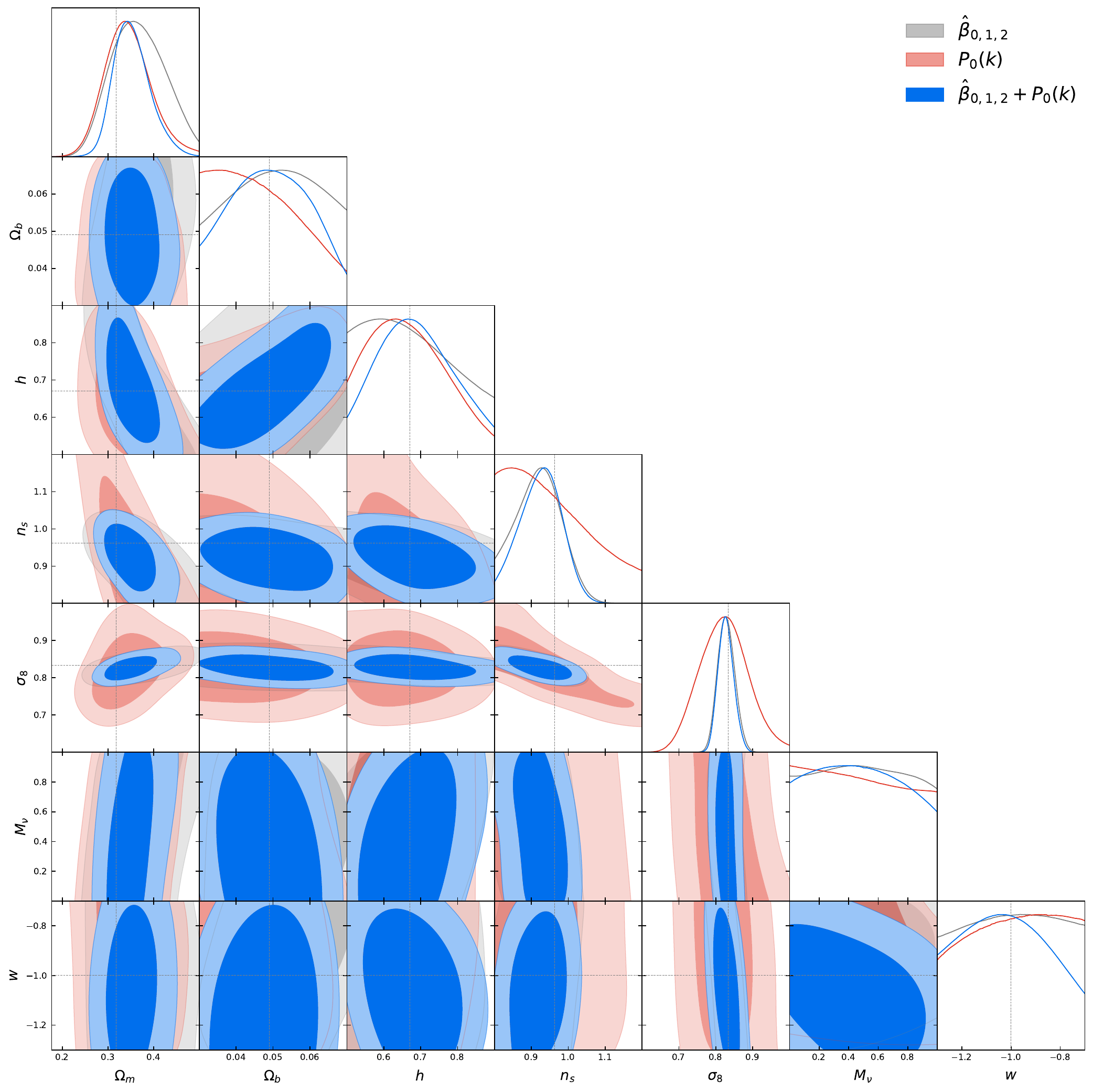}
    \caption{Constraints from Betti curves and power spectrum under fiducial cosmology without RSD.
    The gray, red, and blue contours stand for the constraints from Betti curves, power spectrum, and their combination, respectively.
    The contours mark 68\% (1-$\sigma$) and 95\% (2-$\sigma$) regions of the posteriors. The crossed dashed lines mark the true values for the parameters.}
    \label{fig:BC-Pk_fid}
\end{figure*}
\rev{We compute the power spectrum multipoles in the $k$-range $[0.018,0.3]~h~\mathrm{Mpc}^{-1}$ with bin width $\Delta k = 0.012~h~\mathrm{Mpc}^{-1}$. The power spectrum is modeled using the same AutoML-based emulator framework trained on the identical simulation set as the Betti curve emulator. This ensures that the comparison between the two statistics is performed under comparable modeling complexity and interpolation accuracy. In addition, we have tested EFTofLSS \cite{FOLPs} full-shape models within the \texttt{desilike}\footnote{\url{https://github.com/cosmodesi/desilike}} framework in the $k$-range $[0.018,0.204]~h~\mathrm{Mpc}^{-1}$ \footnote{This $k$-range is because the accuracy of EFT models becomes limited toward the nonlinear regime (e.g., $k \gtrsim 0.2-0.3\ h\ \rm Mpc^{-1}$ \cite{EFT})}, which yielded parameter constraints similar to (in the case of for $\sigma_8,~w,~\text{and }\Omega_{\rm m}$, even weaker than), those from our emulator-based modeling.}
Not considering RSD, the joint cosmological parameter constraints from Betti curves and the monopole of the power spectrum ($P_0(k)$) under the fiducial cosmology are shown in Figure~\ref{fig:BC-Pk_fid}. Compared with the power spectrum alone, Betti curves significantly improve the constraints on $n_{\text{s}}$ and $\sigma_{\text{8}}$, reducing their uncertainties by \rev{46\% and 63\%, respectively.} When combining Betti curves with the power spectrum, \rev{the constraints on $n_{\text{s}}$, $\sigma_{\text{8}}$ are further tightened by 50\%, 69\%, respectively, due to their different parameter-degeneracy directions, while the constraint on $\Omega_{\text{m}}$ and $w$ improves by 19\% and 8\%.} These results demonstrate that Betti curves provide complementary cosmological information to the power spectrum \rev{under matched modeling assumptions}, particularly in probing the primordial power spectrum shape, structure growth, and dark energy evolution.
For $\Omega_{\text{b}}$ and $h$, the joint constraints show no significant improvement over those from the power spectrum alone, indicating that Betti curves are less sensitive to the baryonic composition and the overall spatial scale.
As for $M_{\nu}$, neither Betti curves nor the power spectrum yield meaningful constraints. As discussed in Section~\ref{sec:fid_BC}, this is likely due to systematic uncertainties in the emulator; therefore, we exclude $M_{\nu}$ from the following discussion.

\begin{figure*}
    \includegraphics[width=\linewidth]{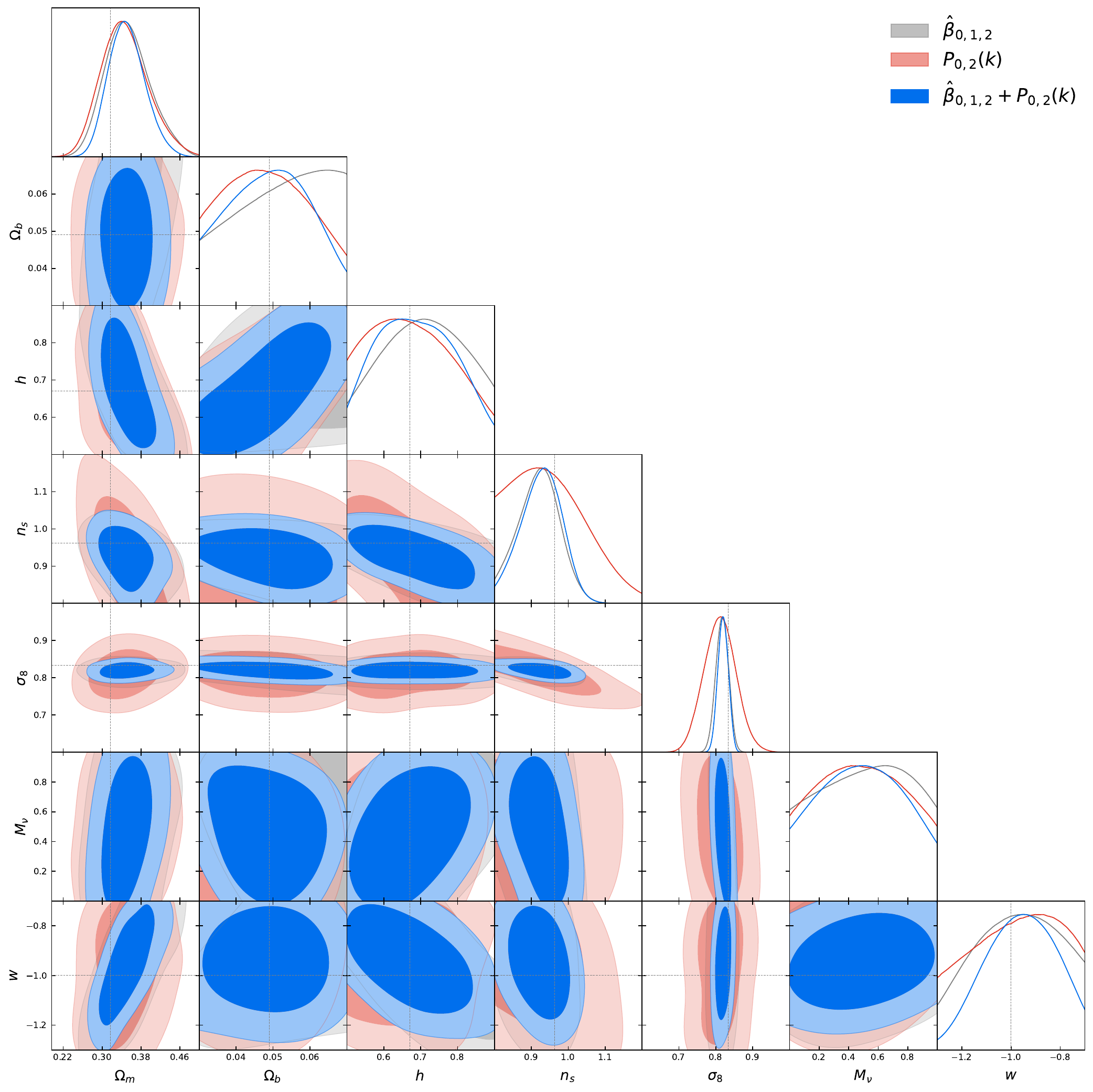}
    \caption{Joint constraints from Betti curves and power spectrum under fiducial cosmology with RSD. Red and blue contours stand for the joint constraints of Betti curves and power spectrum under fiducial cosmology with and RSD. The contours mark 68\% (1-$\sigma$) and 95\% (2-$\sigma$) regions of the posteriors. The crossed dashed lines mark the true values for the parameters.}
    \label{fig:BC-Pk_fid_rsd}
\end{figure*}

Figure~\ref{fig:BC-Pk_fid_rsd} shows the joint constraints from Betti curves and the power spectrum ($P_0(k)$ and $P_2(k)$) under the fiducial cosmology with RSD effects included. \rev{Relative to the power spectrum alone, the joint analysis improves constraints on $n_{\text{s}}$, $\sigma_{\text{8}}$, $\Omega_{\text{m}}$, and $w$ by 43\%, 67\%, 22\%, and 18\%, respectively, while yielding modest improvements of 5\% and 3\% for $\Omega_{\text{b}}$ and $h$. The enhanced complementarity in the RSD case reflects the additional velocity information encoded differently in topological and two-point statistics.}

\subsection{Sub-box result\label{sec:sub result}}
\begin{figure*}
    \includegraphics[width=1\linewidth]{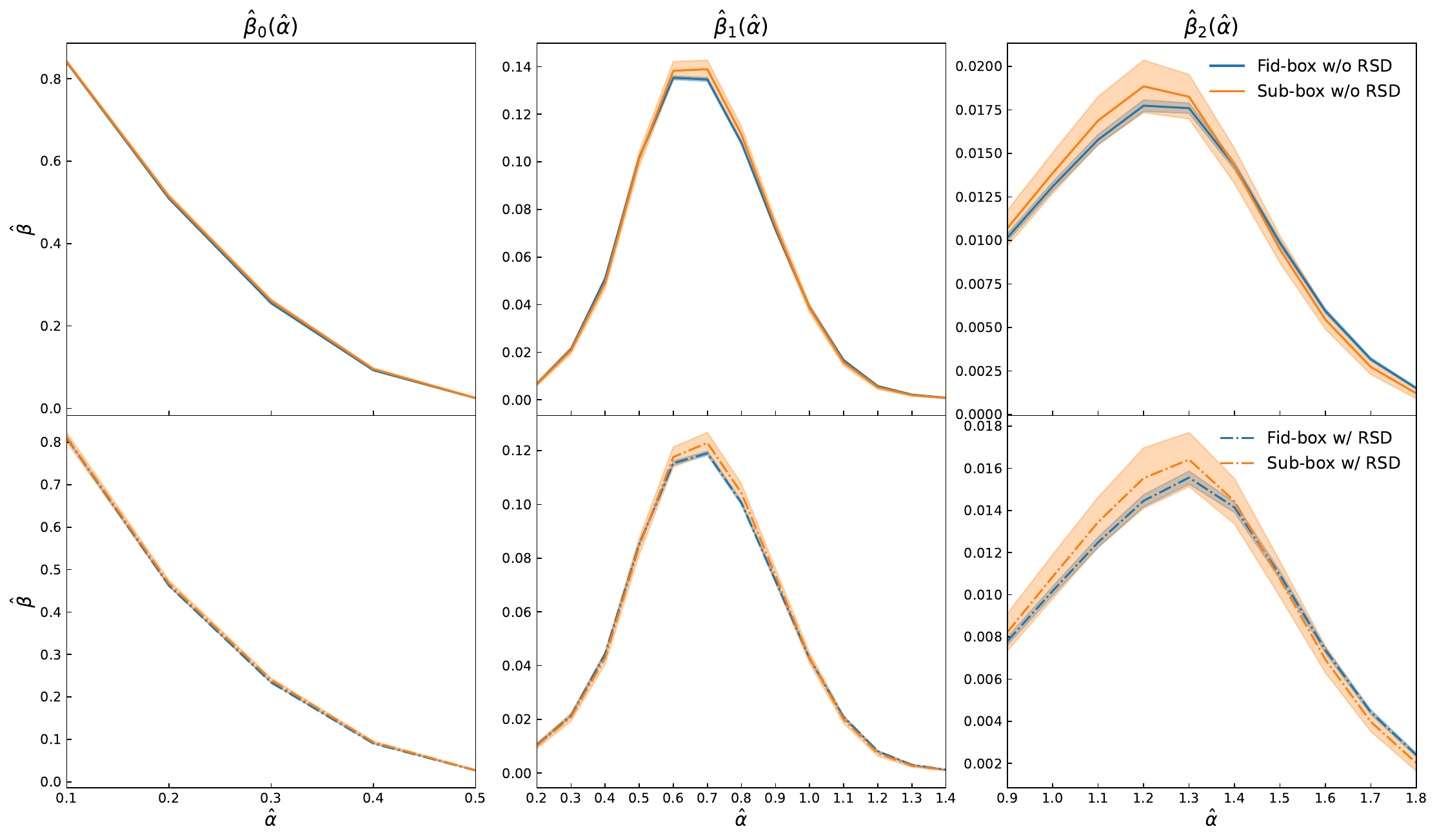}
    \caption{The comparison of Betti curves in fiducial boxes and sub-boxes. The upper panel plots the Betti curves not including RSD effect, lower panel plots the curves including RSD effect. From left to right are $\hat\beta_0$, $\hat\beta_1$, and $\hat\beta_2$. The blue lines with shaded regions stand for Betti curves and error regions in fiducial boxes, while the orange lines and shadow represent Betti curves with errors in sub-boxes.\label{fig:sub vs fid}}
\end{figure*}

We further demonstrate the unbiased nature of parameter constraints obtained from the emulators. Since galaxy surveys observe only a finite portion of the universe, the limited observational volume may not fully capture the statistical properties of the entire universe, an effect known as cosmic variance \cite{CosmicVariance}. To mitigate this, it is advisable to train the emulator on a dataset with a significantly larger volume than the test set. 
To assess whether the emulator predictions generalize to smaller volumes while remaining unbiased, we conduct a sub-box validation test. Since volume is not an explicit input parameter to the emulator, this test evaluates its robustness under varying observational scales. We divide each realization of the validation catalog into sub-boxes, each with a box size of 368 $(\text{Gpc}/h)^3$ (A volume of approximately 1/20 of the fiducial box), and perform parameter recovery tests on the sub-boxes. 
For RSD effect, we first introduce RSD into the fiducial-box simulations using Equation~\eqref{eq:RSD}, and then divide them into sub-boxes.
As described in Section~\ref{sec:normalize}, the computation of Betti curves for the fiducial-box simulations employs 3D periodic boundary conditions.
\rev{Although sub-box division breaks the periodicity of the original simulation volume, we continue to apply periodic boundary conditions when computing Betti curves to preserve a uniform analysis pipeline. This deviation from periodicity also provides an opportunity to test the sensitivity of our framework to boundary assumptions. While the emulator is trained on simulations with periodic boundaries, unbiased parameter recovery on non‑periodic sub‑volumes would suggest that, at the scales considered here, the Betti-curve measurements are not strongly sensitive to boundary assumptions. In other words, the cosmological information captured by Betti curves appears to be dominated by topological structure rather than by features introduced at the box edges.}
Figure~\ref{fig:sub vs fid} compares the Betti curves from fiducial boxes and sub-boxes. Due to the rescaling described in Section~\ref{sec:normalize}, the Betti curves from sub-boxes align well with those from fiducial boxes, though sub-boxes exhibit greater statistical uncertainty due to their smaller volume.

\begin{figure*}
    \includegraphics[width=1\linewidth]{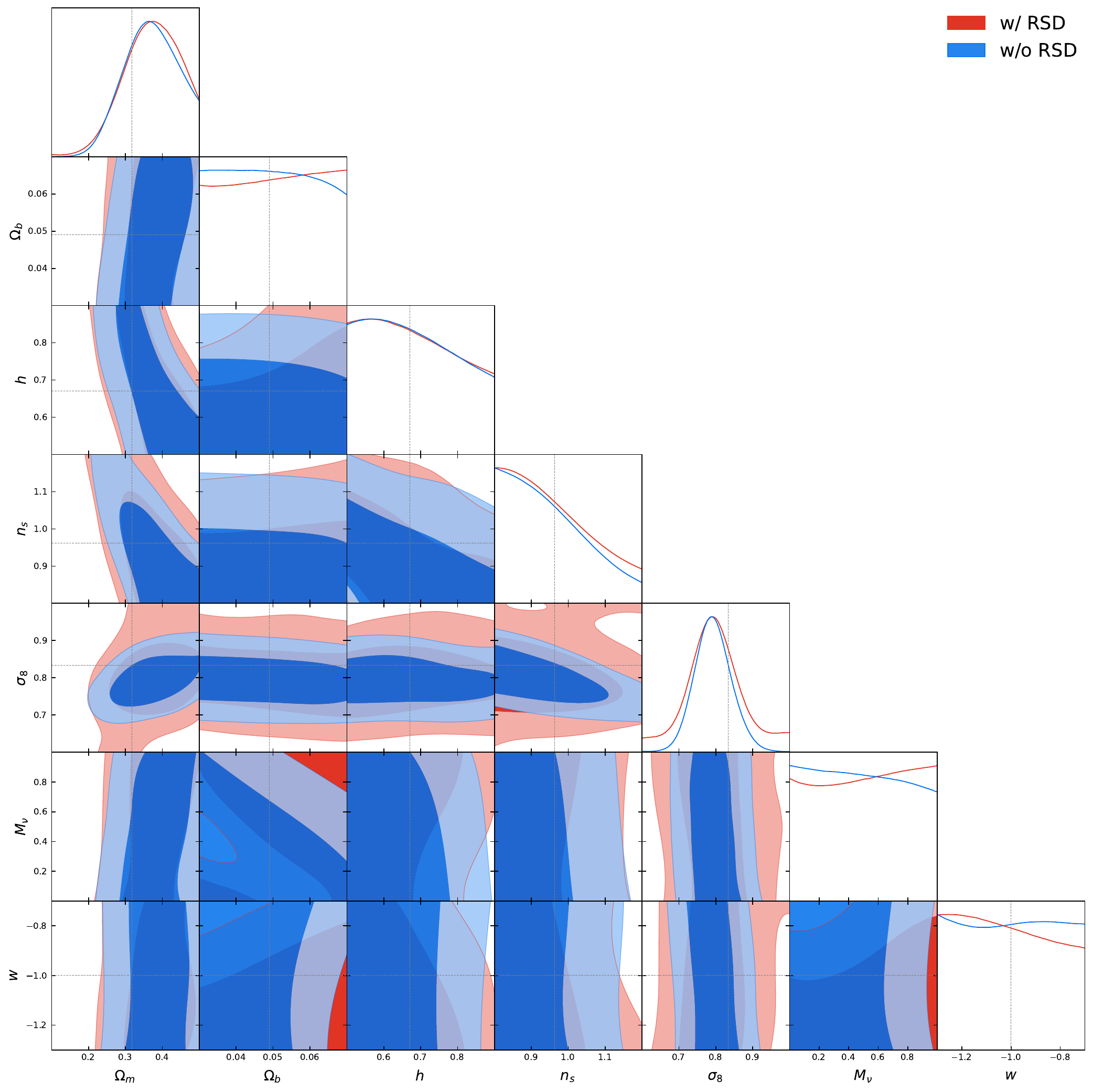}
    \caption{The recovered parameter distributions for fiducial cosmological with or without RSD in sub-boxes. The blue (without RSD) and red (with RSD) contours mark 68\% (1-$\sigma$) and 95\% (2-$\sigma$) regions of the posteriors. The crossed dashed lines mark the true values for the parameters.\label{fig:fza sub post}}
\end{figure*}

We perform parameter inference tests on the sub-box simulations using the same emulators and inference procedure as for the fiducial boxes.
The parameter recovery result, shown in Figure~\ref{fig:fza sub post}, indicates that while the constraints weaken due to the increase in statistical error, the Betti curves continue to constrain $\{\Omega_{\text{m}}, \sigma_8\}$  without bias as concluded in Section~\ref{sec:fid result}.
This demonstrates that the predictive performance of Betti curve emulators is not systematically affected by changes in simulation volume and the break of periodic boundary conditions.
Such robustness is particularly important for real survey analyses, where the observed survey volume is often smaller than that of the training simulations, yet the Betti curves can still provide reliable cosmological parameter estimates.

\section{Conclusions and Discussions \label{sec:conclusions}}

\rev{We developed a coherent and extendable framework on emulator-based cosmological constraints for Betti curves, multiscale topological statistics derived from persistent homology.}
Using dark matter halo catalogs from the \textsc{QUIJOTE} simulations, we develop a complete pipeline to extract topological features from the large-scale structure (LSS) and to constrain cosmological parameters by combining automated machine learning with Bayesian inference.
The proposed framework includes:
\begin{itemize}
\item Periodic $\alpha$-filtration to characterize the cosmic web structure,
\item Scale normalization of Betti curves,
\item Signal-to-noise–driven feature selection,
\item Gaussian process–based emulators optimized via automated machine learning, and
\item Bayesian parameter inference using nested sampling.
\end{itemize}

Based on this pipeline, we investigate the constraining power of Betti curves, their response to RSD, and their joint performance with the power spectrum.
Our key findings are summarized as follows:
\begin{itemize}
\item Among $\hat\beta_0$, $\hat\beta_1$, and $\hat\beta_2$, $\hat\beta_1$ provides the strongest cosmological constraints, achieving the optimal balance between signal-to-noise ratio and parameter sensitivity.
The complementary degeneracy directions of different Betti orders allow their combination to significantly enhance overall parameter constraints.

\item Betti curves show pronounced sensitivity to the spectral index $n_{\text{s}}$, the structure growth amplitude $\sigma_8$, and the matter density parameter $\Omega_{\text{m}}$, achieving constraint precisions of \rev{6.6\%, 2.9\%, and 15.7\%} within a $1\ h^{-3}\mathrm{Gpc}^3$ volume.  
This sensitivity originates from Betti curves’ ability to trace the hierarchical formation of structures: $\sigma_8$ determines the formation strength, $n_{\text{s}}$ the scale distribution, and $\Omega_{\text{m}}$ the overall abundance of structures.  

\item The inclusion of RSD enhances the constraining power on $\sigma_8$, $\Omega_{\text{m}}$, and $w$ by \rev{27\%, 19\%, and 10\%}, respectively.  
This improvement arises because the Fingers-of-God effect provides small-scale velocity dispersion information, while the Kaiser effect introduces large-scale growth constraints.  

\item Betti curves are highly complementary to the power spectrum.  
Their joint analysis breaks degeneracies among $\{\sigma_8, n_{\text{s}},\Omega_{\text{m}}, w\}$ and tightens constraints by \rev{69\%, 50\%, 19\%, and 8\%}, respectively, relative to the power spectrum alone.  
This demonstrates that Betti curves capture cosmological information beyond that contained in traditional two-point statistics.  

\item The RSD effect modifies the degeneracy directions of Betti curve constraints.  
Without RSD, the joint constraints of Betti curves and the power spectrum on $\{\Omega_{\text{b}}, h\}$ are comparable to those from the power spectrum alone.  
When RSD is included, however, the combined constraints improve by \rev{5\% and 3\%,} respectively.

\end{itemize}

Finally, we validate \rev{the assumption of cosmological independence of the covariance and} the robustness of our inference pipeline using sub-box simulations.
The Betti curves retain unbiased constraints on $\sigma_8$, $\Omega_{\text{m}}$, and $n_{\text{s}}$, confirming that our normalization scheme effectively removes volume dependence.
This result demonstrates the generalization ability of the proposed framework across different simulation volumes, laying the groundwork for applying Betti curve to real survey data.

\rev{Despite the promising results for several cosmological parameters, not all parameters are well constrained within the current framework, especially neutrino mass. It implies that Betti curves are not a universal probe capable of capturing all cosmological signatures with equal sensitivity. Therefore, Betti curves should be regarded as complementary statistics that enhance constraints when combined with other statistics. In particular, joint analyses with two-point statistics and beyond-two-point statistics, such as bispectrum, $k$-nearest neighbor distributions, Minkowski functionals, void size functions, and counts-in-cells statistics, will likely be necessary to meet the precision requirements of Stage-V surveys. It is worth to note that the framework developed in this work can be easily extended to such studies, and they will be studied in detail in future work.}

\rev{For practical application to observational data, several additional challenges must also be addressed. Our current pipeline is constructed from halo catalogs and therefore does not incorporate galaxy–halo connection effects. In real surveys, galaxies trace halos in a stochastic and potentially environment-dependent manner, and processes such as halo occupation, assembly bias, and satellite distributions can alter the topology of the observed point distribution. Assessing the robustness of Betti curves under these effects is an essential step toward realistic applications.}
In addition, observational systematics, such as inhomogeneous sampling, survey geometry, and masking effects, can distort Betti curve measurements and introduce biases in cosmological parameter inference.
To mitigate these effects, weighted correction techniques similar to those used in galaxy power spectrum analyses, such as the FKP weighting scheme \cite{FKPweight}, may be required.

Moreover, Betti curves are inevitably affected by shot noise, which mixes stochastic fluctuations with cosmological information.
Although Techniques such as Distance-To-Measure (DTM) \cite{DTM} can reduce shot noise contamination, applying DTM in periodic simulations is computationally expensive, requiring an $\mathcal{O}(N^2)$ distance matrix computation.
A potential alternative is to adapt the random catalogue technique widely used in two-point correlation function estimations \cite{2pcfEstimator1993} to statistically separate the shot-noise contribution from Betti curves.
However, this idea requires further development and testing.

\rev{Overall, this study bridges topological data analysis and cosmology, provides a robust framework for cosmological analysis based on Betti curves, which can be extendable to other beyond-two-point statistics without analytic model.}
It offers a new perspective for studying the formation and evolution of the large-scale structure.
With the advent of upcoming Stage-V surveys, Betti curves are expected to become a valuable complement to standard cosmological probes, enabling more precise and independent parameter constraints.
Future work will focus on extending this framework to real survey data, incorporating realistic observational systematics such as survey geometry and galaxy–halo connection models, \rev{comparing and combining Betti curves with other beyond-two-points statistics such as bispectrum, $k$-nearest neighbor distributions, Minkowski functionals, void size functions, etc.,}
and exploring its application to modified gravity theories, particularly $f(R)$ gravity models.

\begin{acknowledgments}
We thank Yu Liu for useful discussions on this project and Anning Gao for assistance in data access. We acknowledge the support from National Key R\&D Program of China (grant no. 2023YFA1605600).
We acknowledge the Tsinghua Astrophysics High-Performance Computing platform at Tsinghua University for providing computational and data storage resources that have contributed to the research results reported within this paper.
\end{acknowledgments}

\appendix*

\section{Computational Scaling of Betti Curve Measurements}

\rev{To assess the computational scalability of Betti curve measurements, we perform benchmarking using halo catalogs of varying sizes. The full simulation boxes used in this work typically contain $\sim 3\times10^5$ per realization.
For each sub-sampled halo catalog, we measure CPU time and peak Resident Set Size (RSS) during Betti curve computation.
Each configuration is repeated multiple times to estimate the mean and standard deviation.
The tested halo numbers range from $10^3$ to $3\times10^5$. All benchmarks are performed on a machine equipped with Intel(R) Xeon(R) Silver 4210R CPU (2.40 GHz). Each Betti curve computation is executed using a single core.
}

\begin{figure}
    \centering
    \includegraphics[width=\linewidth]{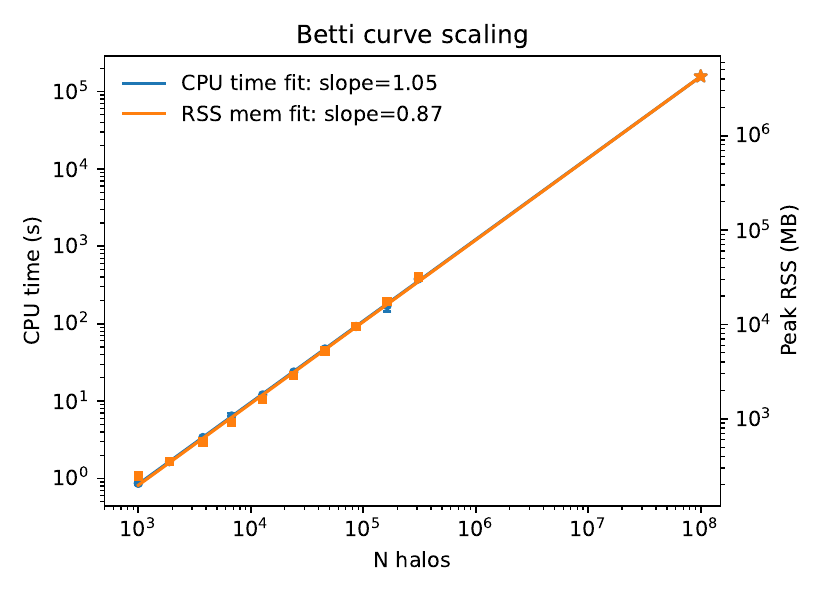}
    \caption{\rev{CPU time and RSS of Betti curve measurements using halo catalogs of varying sizes. The blue circles are the measurements of CPU time and the orange squares are the measurements of RSS. The solid lines are the result of power law fitting. The stars mark the CPU time and RSS extrapolate to Stage-V Survey Volumes.}}
    \label{fig:scaling_plot}
\end{figure}

\rev{Figure~\ref{fig:scaling_plot} shows the scaling behavior of CPU time and peak RSS memory usage as functions of halo number. Both quantities follow approximate power-law relations. A power-law fit yields:
\begin{subequations}\begin{eqnarray}t_{\rm CPU} \propto N^{1.05}_{\rm halo}\\ \rm RSS \propto N_{\rm halo}^{0.87}\end{eqnarray}\end{subequations}
}
\rev{The nearly linear scaling of CPU time indicates that the computational complexity grows approximately proportionally to halo number. The sub-linear memory scaling suggests that memory growth is moderate relative to catalog size.
For the full catalog used in this work ($N_{\mathrm{halo}}\approx 3\times10^5$), the mean CPU time is $t_{\rm CPU} \approx 378\ \rm s$ (single core), and the peak memory usage is approximately $30\ \rm GB$.
}

\rev{These results demonstrate that Betti curve computation remains computationally tractable at the halo counts typical of current large-volume simulations. Future Stage-V surveys (e.g., MUST) will contain effective halo or galaxy counts exceeding $10^8$ in the 2030s \cite{MUST}. Assuming the measured scaling relations continue to hold, extrapolation to $N\sim 10^8$ yields an estimated CPU time of $\sim 46.8$ hours (single core) and peak memory usage of $\sim 4.6$ TB. Such requirements exceed the capacity of typical single-node systems but are compatible with modern high-performance computing environments employing distributed memory and parallel execution. Such requirements would require modern high-performance computing environments or distributed-memory implementations. We emphasize that this extrapolation represents a conservative estimate based on the current implementation. In practice, large survey datasets can be processed using domain decomposition and parallel persistent-homology solvers, substantially reducing both memory pressure and wall-clock time. Continued algorithmic optimization, such as more efficient data structures, filtration pruning, and GPU acceleration, can further improve scalability.
Moreover, computational hardware is expected to advance significantly in the coming decade, with increased memory bandwidth and larger per-node memory capacities. Therefore, while naive scaling suggests substantial resource requirements for Stage-V volumes, no fundamental algorithmic barrier prevents the application of Betti curve statistics to next-generation survey datasets.}

\bibliography{ref}

\end{document}